\documentclass[twocolumn,showpacs,floatfix,preprintnumbers,amsmath,amssymb]{revtex4}
\usepackage{graphicx}
\usepackage{amsmath}
\usepackage{bm}
\usepackage{color}

\newcommand{\bald}[1]{{\bf #1}}

\usepackage{graphicx}
\usepackage{dcolumn}
\usepackage{bm}

\begin{document}


\title{Fast partons as a source of energy and momentum in a perturbative quark-gluon plasma}

\author{R. B. Neufeld}
   \affiliation{Department of Physics, Duke University, Durham, North Carolina 27708, USA}
    \email{rbn2@phy.duke.edu}

\received{3 May 2008}

\begin{abstract}
I derive the space-time distribution of energy and momentum deposited by a fast parton traversing a perturbative quark-gluon plasma by considering the fast parton as the source of an external color field interacting with the medium.  I include the medium's response to screen the fast parton's color field by incorporating dielectric functions and compare to the unscreened result.
\end{abstract}


\maketitle

\section{Introduction}
Quantum chromodynamics (QCD) predicts a transition in sufficiently hot and/or dense nuclear matter from colorless hadrons to a state of deconfined quarks and gluons known as the {\it quark-gluon plasma} (QGP) \cite{Shuryak:1980tp}.  It is believed that this transition has been observed in experiments performed at the Relativistic Heavy-Ion Collider (RHIC) (see, for example, \cite{Adcox:2004mh}).  Fast partons (partons with velocity approaching the speed of light) created by hard transverse scattering in the early moments of a heavy-ion collision are a useful probe in understanding QGP physics.  Of particular importance is the process of {\it jet quenching} in which fast partons lose energy and momentum by interacting with the surrounding medium (see, for instance, \cite{Wang:1991xy,Baier:1996sk,Zakharov:1997uu,Gyulassy:2000fs,Guo:2000nz,Baier:2000mf,Armesto:2004ud,Jacobs:2004qv}).

While much effort has been devoted to understanding the dynamics of jet quenching, the question of how the energy and momentum deposited by the fast parton affects the bulk behavior of an evolving QGP is relatively new (see, for example, \cite{CasalderreySolana:2004qm,Stoecker:2004qu,Satarov:2005mv,Ruppert:2005uz,Renk:2005si,Chaudhuri:2005vc,Friess:2006fk,Casalderrey-Solana:2007km,Chesler:2007sv}).  One promising approach is to treat the fast parton as the source of energy and momentum coupled to the hydrodynamic equations of the medium.  There is strong evidence \cite{Ludlam:2005gx,Gyulassy:2004zy} that the matter created in heavy-ion collisions at RHIC can be described by nearly ideal hydrodynamics and it was argued in \cite{Chaudhuri:2005vc} that if the energy and momentum deposited by a fast parton thermalize quickly the resulting disturbance should evolve hydrodynamically as well.

Casalderrey-Solana {\it et al.} \cite{CasalderreySolana:2004qm} suggested that since the fast parton travels through the medium at supersonic speed, one might expect a Mach cone to develop in the surrounding hydrodynamic medium.  They proceeded to show that if one uses an appropriately chosen supersonic sound source in a linearized hydrodynamic evolution one indeed observes a propagating Mach cone.  In their work, the explicit calculation of the source term generated by a supersonic parton was not addressed.  In a later work \cite{Chaudhuri:2005vc} Chaudhuri and Heinz examined a two-dimensional hydrodynamics simulation of an expanding QGP with a phenomenologically motivated ansatz as their fast parton source term.  Here the authors failed to find the formation of a Mach cone except for seemingly unphysical values of energy loss.  Much recent work \cite{Friess:2006fk,Chesler:2007sv} has been focused on examining a supersonic color charge passing through an ${\cal N}=4$ supersymmetric Yang-Mills plasma (in both of the above citations the authors find a Mach cone in the resulting dynamics).  Such efforts are motivated by the idea that the system is a useful model of a real QCD plasma.

As the above examples illustrate, there have been several insightful attempts at understanding the influence of fast partons on the bulk evolution of an expanding QGP.  However, the explicit evaluation of the distribution of energy and momentum deposited by a supersonic parton traversing a QCD plasma has remained an open question.  It has been pointed out \cite{CasalderreySolana:2004qm,Betz:2008js} that the medium's response to a fast parton is sensitive to the specific form of energy and momentum deposition.  For example, the hydrodynamic source vector of a parton moving at constant velocity, $\bald{u}$, can be written in Fourier space as \cite{Chesler:2007sv}
\begin{equation}
\bald{J}(\omega, \bald{q}) = \left[\bald{u} \, \phi_u(\omega, q^2) + i \bald{q} \phi_q(\omega, q^2)\right]2 \pi \delta(\omega - \bald{u}\cdot \bald{q}).
\end{equation}
The medium's response then becomes sensitive to the functions $\phi_u$ and $\phi_q$.  For instance, if one sets $\phi_u = 0$ then the source fails to excite diffusive momentum density in the medium.  The full momentum dependence of the source is necessary to predict the spectrum and shape of the sound wave excited in the medium.  The sensitivity of the medium's response to the specific mechanism of energy and momentum deposition creates the need for a hydrodynamic source term derived from first principles.

In order to derive such a source term one must first understand the mechanism of energy and momentum deposition and the different scales involved.  Consider a weakly coupled QCD plasma at an asymptotically high temperature $T$.  In this system a fast parton interacts with the medium and creates a disturbance at a distance scale of the order $(g T)^{-1}$, where $g$ is the running coupling.  The disturbance in turn interacts with the surrounding medium, creating a new disturbance which again interacts with the medium, and the process continues, each time at larger and larger distance scales.  One could, in principle, determine the medium's response to a fast parton at arbitrarily large distances by calculating each subsequent disturbance up to the desired scale.  In practice, however, it is desirable to use an effective theory, which becomes valid at some length scale, to model the medium's response at long distances.  As mentioned previously, it is believed that nearly ideal hydrodynamics is an accurate model of the long distance behavior of the QGP produced at RHIC.  Hydrodynamics becomes valid at length scales which are large compared to the mean free path.  Again, considering a weakly coupled QCD plasma at high temperature, $T$, the transport mean free path is of order $(g^4 T)^{-1}$ \cite{Arnold:2002zm}.  Thus to derive an effective QCD hydrodynamic source term one should calculate the medium's response up to distances (at least) of order $(g^4 T)^{-1}$, at which point the system evolves hydrodynamically.

Recently, Neufeld {\it et al.} \cite{Neufeld:2008fi} gave a schematic derivation of the hydrodynamic source term expected from a fast parton moving through a QGP in the relativistic limit (velocity approaching the speed of light) in the absence of in-medium screening by including the medium response at a distance scale of the order $(g T)^{-1}$.  The authors found a propagating Mach cone in the resulting linearized hydrodynamic evolution.  In this work, I will explicitly derive this distribution in the context of kinetic theory including color screening.  The approach will be to treat the fast parton as the source of an external color field interacting with a perturbative QGP through the collisionless Boltzmann equation.  After obtaining the microscopic equation of motion, I find the macroscopic evolution by taking moments via the usual Chapman-Enskog approach and making the assumption of local thermal equilibrium.  I find that the fast parton serves as a source distribution of energy and momentum for the hydrodynamic equations of the medium.  I then perform an explicit evaluation of the distribution both with and without including the medium's response to screen the fast parton's color field and interpret the results.

As mentioned above, a consistent derivation of the hydrodynamic source term should include medium response to distances (at least) of order $(g^4 T)^{-1}$.  In that sense the distribution derived here can be thought of as a sophisticated representation of a delta function.  The calculation of the source at intermediate scales will be presented in a future publication (in what follows, I choose units such that $\hbar = c = k_b = 1$).

\section{Hydrodynamic Source Term}
Consider a system of partons in the presence of an external color field $A_\mu^a$ and described by the distribution $f({\bf x},{\bf p},t,Q)$ which includes the color sector through its $Q$ dependence.  The Boltzmann equation for this distribution is given by
\begin{equation}\label{vbstart}
\left(\frac{\partial}{\partial t} + \frac{{\bf p}}{E}\cdot {\bm\nabla}_x + \frac{{d \bf p}}{d t}\cdot {\bm\nabla}_p + \frac{\partial Q^a}{\partial t}\frac{\partial}{\partial Q^a}\right)f({\bf x},{\bf p},t,Q) = 0
\end{equation}
where the collision term is omitted.  The chromodynamic equations of motion for a parton with charge $g Q^a(t)$ in a color field $A_\mu^a$ are given by \cite{Wong:1970fu}
\begin{equation}\label{qcd1}
\frac{{d \bf p}}{d t} = g {Q^{a}(t)}\left({\bf E}^a + ({{\bf v}\times{\bf B}})^{a}\right)
\end{equation}
and
\begin{equation}\label{qcd2}
\frac{{d Q^a(t)}}{d t} = - g f_{abc}A_\mu^b v^\mu Q^c(t)
\end{equation}
where $v_\mu = d x_\mu/d t$, $f_{abc}$ are the $SU(3)$ structure numbers, and
\begin{equation}
{\bf E}^a + ({{\bf v}\times{\bf B}})^{a} \equiv {\bf F}^a
\end{equation}
is the color Lorentz force.  Inserting (\ref{qcd1}) and (\ref{qcd2}) into (\ref{vbstart}) yields
\begin{equation}\label{vbmore}
\begin{split}
\left(v^\mu \frac{\partial}{\partial x^\mu} + g {Q^{a}(t)}{\bf F}^a\cdot {\bm\nabla}_p - g f_{abc}A_\mu^b v^\mu Q^c(t)\frac{\partial}{\partial Q^a}\right)& \\
\times f({\bf x},{\bf p},t,Q)= &0.
\end{split}
\end{equation}

The singlet and octect components of the parton distribution, $f({\bf x},{\bf p},t)$ and $f^a({\bf x},{\bf p},t)$, are obtained by taking the moments of $f({\bf x},{\bf p},t,Q)$ in the color sector:
\begin{equation}\label{mom1}
f({\bf x},{\bf p},t) = \int d Q\, f({\bf x},{\bf p},t,Q) \equiv f_0
\end{equation}
and
\begin{equation}\label{mom2}
f^a({\bf x},{\bf p},t) = \int d Q\, Q^a f({\bf x},{\bf p},t,Q) \equiv f_1^a.
\end{equation}
The notation $f_1^a$ used in (\ref{mom2}) is meant to emphasize that any contribution from the color octet distribution must come from off equilibrium effects since it vanishes in equilibrium.  Assuming $g A_\mu^a$ can be treated as a perturbation in an otherwise equilibrated system the contribution from $f_1^a$ should be small and higher moments, such as $f_2^{ab}$ ($a \neq b$), which are of higher order in $g A_\mu^a$, will be ignored.  Applying the first two moments to (\ref{vbmore}) gives
\begin{equation}\label{vbprog}
\begin{split}
v^\mu \frac{\partial f_0}{\partial x^\mu} + g {\bf F}^a\cdot {\bm\nabla}_p f_1^a= 0
\end{split}
\end{equation}
and
\begin{equation}\label{vbgauge}
\begin{split}
v^\mu D_\mu f_1^a = - \frac{g C_2}{N_c^2 - 1} {\bf F}^a\cdot {\bm\nabla}_p f_0
\end{split}
\end{equation}
where $D_\mu f_1^a = \partial_\mu f_1^a + g f_{abc}A_\mu^b f_1^c$ is the covariant derivative.  In obtaining (\ref{vbgauge}) I have used
\begin{equation}
\int d Q\, Q^a Q^b f({\bf x},{\bf p},t,Q) = \frac{C_2 \delta^{ab}}{N_c^2 - 1}f_0 + f_2^{ab}
\end{equation}
where $C_2$ is the eigenvalue of the Casimir operator of the medium partons and, as mentioned above, $f_2^{ab}$ is neglected.

An equation for $f_0$ can be obtained by solving (\ref{vbgauge}) for $f_1^a$ and inserting the result into (\ref{vbprog}).  Neglecting any space-time dependence in $f_0$ the solution is given by
\begin{equation}
\begin{split}\label{primefone}
f_1^a = - \frac{i g C_2}{N_c^2 - 1} \int \frac{d^4 k}{(2 \pi)^4} \int d^4 x' U_{ab}(x,x')\times\\
\frac{e^{i k \cdot (x' - x)}}{v \cdot k + i \epsilon}{\bf F}^b(x')\cdot {\bm\nabla}_p f_0
\end{split}
\end{equation}
where
\begin{equation}\label{comp}
U_{ab}(x,x') = P \exp {\left(-\int_{x'}^{x}g f_{acb} A_\mu^c d x^\mu \right)}
\end{equation}
is the path ordered gauge connection.  The result ({\ref{primefone}}) can be simplified with a contour integration in $k^0$.  It has one pole which is in the lower complex plane (i.e., $t'<t$).  I find
\begin{equation}
\begin{split}\label{nice}
f_1^a = - \frac{g C_2}{N_c^2 - 1} \int \frac{d^3 {\bf k}}{(2 \pi)^3} \int d^4 x' U_{ab}(x,x') \\
\times e^{i {\bf v}\cdot{\bf k}(t' - t) - i {\bf k} \cdot ({\bf x'} - {\bf x})} {\bf F}^b(x')\cdot {\bm\nabla}_p f_0 \\
= - \frac{g C_2}{N_c^2 - 1} \int_{-\infty}^t d t' U_{ab}(x,x') {\bf F}^b(x')\cdot {\bm\nabla}_p f_0.
\end{split}
\end{equation}
where now ${\bf x'} = {\bf x}(t') = {\bf v}(t' - t) + {\bf x}$.  Combining (\ref{nice}) with (\ref{vbprog}) gives the result for $f_0$ \cite{Asakawa:2006jn}
\begin{equation}\label{mainvb}
v^\mu \frac{\partial f_0}{\partial x^\mu} - \nabla_{p_i} {D_{ij}}\nabla_{p_j}f_0 = 0
\end{equation}
where
\begin{equation}\label{dupree}
{D_{ij}} = \frac{g^2 C_2}{N_c^2 - 1}\int_{-\infty}^t d t' F_i^a(x) U_{ab}(x,x') F_j^b(x').
\end{equation}

The result (\ref{mainvb}) describes the phase space distribution of partons in the presence of a soft external color field $A_\mu^a$ (the term {\it soft} implies the momenta in $A_\mu^a$ are small compared to the average momentum in $f_0$).  My approach is to consider the external color field to be generated by a fast parton propagating through the medium, which I specify to be a perturbative QGP.  In light of the perturbative nature of the medium it makes sense to expand the path ordered gauge connection in (\ref{dupree}) to zeroth order in $g$.  In the hard thermal loop (HTL) approximation the source field for a fast parton with constant velocity ${\bf u}$ at position ${\bf x} = {\bf u}t$ has the Fourier representation
\begin{equation}
\label{efieldT}
\bald{E}^a_T(x) = \frac{i g {Q_p^a}}{(2 \pi)^3}\int d^4k \,e^{-ik\cdot x} \frac{\left(\bald{u}k^2 - \bald{k}(\bald{k}\cdot\bald{u})\right)\omega\delta(\omega - \bald{u}\cdot\bald{k})}{k^2(k^2 - \epsilon_T(\bald{k},\omega)\omega^2)}
\end{equation}
\begin{equation}\label{efieldL}
\begin{split}
\bald{E}^a_L(x) = -\frac{i g {Q_p^a}}{(2 \pi)^3}\int d^4k \,e^{-ik\cdot x}\frac{(\bald{k}\cdot\bald{u})\delta(\omega - \bald{u}\cdot\bald{k})}{\epsilon_L(\bald{k},\omega) \omega k^2}\bald{k}
\end{split}
\end{equation}
\begin{equation}
\label{bfield}
\bald{B}^a(x) = \frac{i g {Q_p^a}}{(2 \pi)^3}\int d^4k \,e^{-ik\cdot x}\frac{(\bald{k}\times\bald{u})\delta(\omega - \bald{u}\cdot\bald{k})}{k^2 - \epsilon_T(\bald{k},\omega)\omega^2}
\end{equation}
where $k^\mu = (\omega,{\bf k})$, and the dielectric functions, $\epsilon_L(\bald{k},\omega)$ and $\epsilon_T(\bald{k},\omega)$, encode the medium's response to screen the fields.  These functions read explicitly \cite{Ruppert:2005uz}
\begin{equation}\label{ep}
\begin{split}
\epsilon_L = 1 + \frac{2 m_g^2}{k^2}&\left[1 - \frac{\omega}{2 k}\left(\ln{\left[\frac{k + \omega}{k - \omega}\right]} - i \pi\Theta(k^2 - \omega^2)\right)\right], \\
\epsilon_T = 1 - \frac{m_g^2}{\omega^2}&\left[\frac{\omega^2}{k^2} + \frac{\omega}{2 k}\frac{(k^2-\omega^2)}{k^2} \right.\\
&\left. \times \left(\ln{\left[\frac{k + \omega}{k - \omega}\right]} - i \pi\Theta(k^2 - \omega^2)\right)\right]
\end{split}
\end{equation}
where $m_g = m_D/\sqrt{3}$ and $\Theta(k^2 - \omega^2)$ is a step function which is unity if $k^2 > \omega^2$ and zero otherwise.  The derivation and momentum space representation of Eqs. (\ref{efieldT} - \ref{bfield}) is given in Appendix A.  Note that in (\ref{efieldT}) and (\ref{efieldL}) the electric field has been divided into transverse and longitudinal parts such that $ {\bf E} = {\bf E}_T + {\bf E}_L$.

The macroscopic equations of motion are found by taking moments of the evolution equation (\ref{mainvb}):
\begin{equation}\label{moment}
\begin{split}
\int\frac{d {\bf p} \,p^\nu}{(2 \pi)^3} \left(v^\mu \frac{\partial f_0}{\partial x^\mu} - \nabla_{p_i} {D_{ij}}\nabla_{p_j}f_0\right) = 0
\end{split}
\end{equation}
where $p^\nu = (E,{\bf p})$.  The integrals in (\ref{moment}) can be evaluated by boosting to the frame comoving with the volume element and then exploiting the hydrodynamic assumption of local thermal equilibrium by using the thermodynamic relations (which are only valid in the local frame of the medium element)
\begin{equation}\label{pressure}
\int\frac{d {\bf p}}{(2 \pi)^3} \frac{p^\mu p^\nu}{p^0}f_0 = (\epsilon + p)\delta^{0 \mu}\delta^{0\nu} - p g^{\mu \nu}
\end{equation}
where $p$ is the local pressure, $\epsilon$ is the local energy density and $g^{\mu \nu}$ is the metric tensor.  Using the notation
\begin{equation}\label{sourceterm}
\int\frac{d {\bf p} \,p^\nu}{(2 \pi)^3} (\nabla_{p_i} {D_{ij}} \nabla_{p_j}f) \equiv J^{\nu}
\end{equation}
I find that the resulting equations of motion for the medium evolution are
\begin{equation}\label{sourcehydro}
\partial_\mu T^{\mu \nu} = J^{\nu}
\end{equation}
where the energy-momentum tensor is given by
\begin{equation}\label{basictensor}
T^{\mu \nu} = (\epsilon + p)w^\mu w^\nu - p g^{\mu \nu}
\end{equation}
and $w^\mu$ is the medium 4-velocity.

The result (\ref{sourcehydro}) describes the macroscopic evolution of a perturbative QGP in the presence of $J^\nu$, which acts as a source of energy and momentum.  In other words, $J^\nu$ gives the distribution of energy and momentum deposited into a perturbative QGP due to the passage of a fast parton.  This is precisely the quantity I set out to evaluate.  In the remainder of this work, I will evaluate the source term (\ref{sourceterm}) both with and without dielectric screening and examine the resulting distribution.

\section{Explicit Evaluation of the Source Term}

I begin by simplifying (\ref{sourceterm}) with an integration by parts
\begin{equation}\label{partsreduce}
\begin{split}
\int\frac{d {\bf p} \,p^\nu}{(2 \pi)^3} &(\nabla_{p_i} {D_{ij}}\nabla_{p_j}f_0) = \\
&-\int\frac{d {\bf p} \,}{(2 \pi)^3} (\delta_{0\nu}\frac{p_i}{p_0} + \delta_{i\nu}){D_{ij}}\nabla_{p_j}f_0
\end{split}
\end{equation}
where $\delta_{0\nu}$ is the Kronecker-Delta symbol and, as mentioned previously, $i$, $j$ range from 1 to 3.  I specify the medium to be a locally thermal plasma of massless gluons with the distribution
\begin{equation}\label{boseeinstein}
f_0 = \frac{2 ({N^{2}_c}-1)}{e^{\beta p^0}-1}
\end{equation}
where $N_c$ is the number of colors and $1/\beta = T$ is the local temperature of the medium.  With this specification the distribution, $f_0$, now contains the only dependence upon the magnitude of the momentum, $p^0 = p$, in (\ref{partsreduce}).  It follows that
\begin{equation}\label{further}
\begin{split}
-\int\frac{d {\bf p} \,}{(2 \pi)^3} (\delta_{0\nu}\frac{p_i}{p_0}& + \delta_{i\nu}){D_{ij}}\nabla_{p_j}f_0 = \\
&\frac{(N_c^2 - 1) T^2}{3}\int\frac{d \hat{{\bf v}}\,\hat{{\bf v}_j}}{4 \pi}(\delta_{0\nu}\hat{{\bf v}_i} + \delta_{i\nu}){D_{ij}}
\end{split}
\end{equation}
where $\hat{{\bf v}} = \bald{p}/p^0$ is the directional velocity of medium particles.

Inserting the explicit form of $D_{ij}$ gives
\begin{equation}\label{jnuadvanced}
\begin{split}
J^{\nu}(x) = &\frac{m_D^2}{(2 \pi)^8} \int_{-\infty}^{t}dt' \int d^4k d^4k' d \hat{{\bf v}} \frac{\hat{{\bf v}_j}}{4 \pi}\times \\
&e^{-i k \cdot x - i k'\cdot x'} (\delta_{0\nu}\hat{{\bf v}_i} + \delta_{i\nu})F^a_i(k)F^a_j(k')
\end{split}
\end{equation}
where $m_{\rm D} = \sqrt{N_c/3}\, gT$ is equal to the Debye mass for gluons in the HTL approximation, and I have introduced the Fourier representation of the color Lorentz force
\begin{equation}
F^a_i(x) = \frac{1}{(2 \pi)^4}\int d^4k \,e^{-ik\cdot x}F^a_i(k).
\end{equation}
The generalization to include $N_f$ flavors of quarks/antiquarks in the medium can be done in a straightforward way by adding the appropriate thermal distribution to (\ref{boseeinstein})
\begin{equation}\label{fulldist}
f_0 = \frac{2 ({N^{2}_c}-1)}{e^{\beta p^0}-1} + \frac{4 N_c N_f}{e^{\beta p^0} + 1}
\end{equation}
and repeating the steps which lead to (\ref{jnuadvanced}).  The only modification is that the expression for the Debye mass appearing in (\ref{jnuadvanced}) now includes flavor: $m_{\rm D} = \sqrt{(2 N_c + N_f)/6}\, gT$.  In the numerical results that follow I have set $N_f = 0$.

I next make the assumption that the typical momentum transfer between the source particle and the medium is small compared to the typical momentum of a medium particle.  This assumption allows a straight line approximation relating the position of a medium particle at time $t'$ to its position at time $t$:
\begin{equation}\label{straight}
{\bf x'} = {\bf x} + \hat{{\bf v}}(t'-t).
\end{equation}
The $t'$ integration can be evaluated after plugging (\ref{straight}) into (\ref{jnuadvanced}) and introducing an exponential damping factor, $i \epsilon$, yielding
\begin{widetext}
\begin{equation}
\begin{split}\label{jnuadv}
J^{\nu}(x) = \frac{i m_{\rm D}^2}{(2 \pi)^8} \int d^4k d^4k'd \hat{{\bf v}} e^{i ({\bf k} + {\bf k'})\cdot {\bf x} - i t (\omega + \omega')} \times \frac{\hat{{\bf v}}\cdot {\bf E}^a(k')\left(\delta_{0\nu}\hat{{\bf v}}\cdot{\bf E}^{a}(k) + \delta_{i\nu} \left({E^{a}_i}(k) + ({\hat{{\bf v}}\times{\bf B}})^{a}_i(k)\right)\right)}{4 \pi(\omega'-{\bf k'}\cdot\hat{ \bf v} + i\epsilon)}.
\end{split}
\end{equation}
\end{widetext}

Next consider the integral over $\hat{ \bf v}$.  It can be evaluated by choosing a frame in which ${\bf k'} = k'\hat{ \bf z}$, performing the integral, and then rotating back into a frame in which ${\bf k'}$ is arbitrary.  The explicit details of the calculation are given in Appendix B.  The result is most easily expressed by dividing the source term into two pieces, $J_{\bf 1}^{\nu}$ and $J_{\bf 2}^{\nu}$, where $J_{\bf 1}^{\nu} + J_{\bf 2}^{\nu} = J^{\nu}$.  After integrating out the delta functions in (\ref{efieldT} - \ref{bfield}) I find
\begin{equation}
\begin{split}\label{jnuone}
J_{\bf 1}^{\nu}(x) = &\frac{i m_{\rm D}^2}{(2 \pi)^8} \int d^3k d^3k'e^{i ({\bf k} + {\bf k'})\cdot ({\bf x} - {\bf u} t )} \Omega_2({\bf k}')\left( \delta_{0\nu} {\bf E}^a({\bf k}') \right. \\
&\left.\cdot{\bf E}^{a}({\bf k}) + \delta_{i\nu}({\bf E}^a({\bf k}')\times{\bf B}^{a}({\bf k}))_i\right)
\end{split}
\end{equation}
and
\begin{equation}
\begin{split}\label{jnutwo}
J_{\bf 2}^{\nu}(x) = &\frac{i m_{\rm D}^2}{(2 \pi)^8} \int d^3k d^3k'e^{i ({\bf k} + {\bf k'})\cdot ({\bf x} - {\bf u} t )} \\
& \times({\bf E}^a({\bf k}')\cdot\hat{\bf k'})\left(\left(\Omega_1({\bf k'}){\bf u}\cdot\hat{\bf k}' - \Omega_2({\bf k'})\right) \right. \\
&\left.\left(\delta_{0\nu}({\bf E}^a({\bf k})\cdot{\bf {\bf k}'}) + \delta_{i\nu}(\hat{\bf k'}\times{\bf B}^{a}({\bf k}))_i\right) \right.\\
&\left. + \delta_{i\nu}\Omega_1({\bf k}'){E^{a}_i}({\bf k})\right)
\end{split}
\end{equation}
where
\begin{equation}\label{g1}
\Omega_1({\bf k}') = \frac{{\bf u}\cdot{\bf k}'}{2 k'^2}\ln{\left[\frac{k' + {\bf u}\cdot{\bf k}'}{k' - {\bf u}\cdot{\bf k}'}\right]}-\frac{i \pi {\bf u}\cdot{\bf k}'}{2 k'^2} -\frac{1}{k'}
\end{equation}
\begin{equation}\label{g2}
\begin{split}
\Omega_2({\bf k}') = &\frac{1}{4 k'^3}\left(\ln{\left[\frac{k' + {\bf u}\cdot{\bf k}'}{k' - {\bf u}\cdot{\bf k}'}\right]}(k'^2 - ({\bf u}\cdot{\bf k}')^2) \right.\\
&\left.-\pi i(k'^2 - ({\bf u}\cdot{\bf k}')^2) + 2 k'{\bf u}\cdot{\bf k}'\right).
\end{split}
\end{equation}
In the above expressions $\hat{\bf k'}$ is the unit vector in the ${\bf k'}$ direction, $k'$ is the magnitude of ${\bf k'}$, ${\bf u}$ is the source parton's velocity.  The exponential damping factor in (\ref{jnuadv}), $i \epsilon$, sets the sign of the imaginary terms in ($\ref{g1}$), ($\ref{g2})$.

At this point it is convenient to specify a direction for the source particle's velocity which I do by choosing ${\bf u} = u \,\hat{\bf z}$.  I also choose to work in plane polar coordinates, $k_T$ and $\phi$, such that $k_x = k_T \cos\phi$ and $k_y = k_T \sin\phi$.  With these choices the only dependence upon the momentum space variable $\phi$ is in the exponential and terms proportional to $\cos\phi$ or $\sin\phi$.  The exponential depends upon $\phi$ through the term $i k_T (x \cos{\phi} + y \sin{\phi})$ which can be rewritten as $i k_T \rho \cos{[\phi - \alpha]}$, where $x = \rho \cos \alpha$ and $y = \rho \sin \alpha$.  With these simplifications the entire $\phi$ dependence of (\ref{jnuone}) and (\ref{jnutwo}) can be integrated out by using the relations
\begin{equation}
\begin{split}
\int_0^{2 \pi} &\frac{d \phi}{2 \pi} \exp{[i k_T \rho (\cos{[\phi-\alpha]})]}
\begin{bmatrix}
1\\
\cos\phi\\
\sin\phi
\end{bmatrix} \\
\equiv
&\begin{bmatrix}
J_0(\rho k_T)\\
\frac{i x}{\rho} J_1(\rho k_T)\\
\frac{i y}{\rho} J_1(\rho k_T)
\end{bmatrix}
\end{split}
\end{equation}
where $J_n(x)$ is a Bessel function of order $n$.  The calculation is further simplified due to the fact that the ${\bf k}$ and ${\bf k}'$ dependence can be completely factorized, reducing the remaining four-dimensional integration to a product of two-dimensional integrations.  Using the explicit field dependence listed above, I find there are 12 unique two-dimensional integrations which must be performed.  They are given by
\begin{equation}\label{kye}
\begin{split}
\begin{bmatrix}
\xi_1\\
\xi_2\\
\xi_3\\
\xi_4
\end{bmatrix}
&\equiv
\int dk_z dk_T \frac{e^{i (z-u t) k_z} \Omega_1(k) }{k D_L(k)}
\begin{bmatrix}
J_0(\rho k_T)k_T k^2 \\
J_0(\rho k_T) k_T k_z^2\\
J_0(\rho k_T) k_T^3\\
J_1(\rho k_T) k_z k_T^2
\end{bmatrix}
\\
\begin{bmatrix}
\xi_5\\
\xi_6
\end{bmatrix}
&\equiv
\int dk_z dk_T \frac{e^{i (z-u t) k_z} \Omega_2(k) }{k^2 D_T(k)}
\begin{bmatrix}
J_0(\rho k_T)k_z k_T^3\\
J_1(\rho k_T) k_T^2 k_z^2
\end{bmatrix}
\\
\begin{bmatrix}
\xi_7\\
\xi_8
\end{bmatrix}
&\equiv
\int dk_z dk_T \frac{e^{i (z-u t) k_z}}{D_L(k)}
\begin{bmatrix}
J_1(\rho k_T)k_T^2\\
J_0(\rho k_T) k_z k_T
\end{bmatrix}
\\
\begin{bmatrix}
\xi_9\\
\xi_{10}\\
\xi_{11}\\
\xi_{12}
\end{bmatrix}
&\equiv
\int dk_z dk_T \frac{e^{i (z-u t) k_z}}{k^2 D_T(k)}
\begin{bmatrix}
J_1(\rho k_T) k_T^2 k^2\\
J_1(\rho k_T) k_T^4\\
J_1(\rho k_T) k_T^2 k_z^2\\
J_0(\rho k_T) k_z k_T^3
\end{bmatrix}
\end{split}
\end{equation}
where $D_L(k) \equiv k^2{\epsilon_L(k)}$, $D_T(k) \equiv k^2 - u^2 k_z^2\epsilon_T(k)$.

In terms of the above definitions the source term is
\begin{equation}
\begin{split}\label{jperp}
J^{x(y)} = &\frac{g^2 {m^{2}_D}({Q_p^a})^2}{16 \pi^4}\frac{x(y)}{\rho} \\
&\times \left(u^3 \xi_5 \xi_9 + ( \xi_1 \xi_7 - u^2 \xi_2 \xi_{10} + u^2 \xi_3 \xi_{11})\right),
\end{split}
\end{equation}
\begin{equation}
\begin{split}\label{jz}
J^{z} &= \frac{i g^2 {m^{2}_D}({Q_p^a})^2}{16 \pi^4}\left(u^2 \xi_9(u \xi_6  + \xi_4) - \xi_1 (\xi_8 - u^2 \xi_{12})\right)
\end{split}
\end{equation}
and
\begin{equation}
\begin{split}\label{jnot}
J^0 = &-\frac{i g^2 {m^{2}_D}({Q_p^a})^2}{16 \pi^4}\left((\xi_8 - u^2 \xi_{12})(u \xi_2 - u^2 \xi_5 ) \right. \\
&\left. - u (u \xi_6  + \xi_4)(\xi_7 + u^2\xi_{11})\right).
\end{split}
\end{equation}
The problem is now reduced to the evaluation of the 12 $\xi_i$ terms.  Before considering the full results of such an evaluation it is useful to calculate the above expressions without including the medium's response to screen the fields.  I perform this calculation in the next section and obtain an analytical result.

\section{The Unscreened Source Term}\label{unscreen}
The unscreened source term is found by setting the dielectric function given in (\ref{ep}) to unity, or, equivalently, setting $D_L(k) = k^2$ and $D_T(k) = k^2 - u^2 k_z^2$ in (\ref{kye}).  It is easy to verify that with this simplification (\ref{jperp}) and (\ref{jnot}) can be rewritten as
\begin{equation}
\begin{split}\label{jperpmod}
J^{x(y)} &= \frac{g^2 {m^{2}_D}({Q_p^a})^2}{16 \pi^4}\frac{x(y)}{\rho}\xi_9\left(u^3 \xi_5  + ( \xi_1 - u^2 \xi_2)\right),
\end{split}
\end{equation}
\begin{equation}\label{sourcenot}
\begin{split}
J^0 = &\frac{i g^2 {m^{2}_D}({Q_p^a})^2}{16 \pi^4} \times \left(u \xi_9(u \xi_6  + \xi_4) \right. \\
&\left.- ( u \xi_2 - u^2 \xi_5 )(\xi_8 - u^2 \xi_{12})\right)
\end{split}
\end{equation}
A quick perusal of the source term given by (\ref{jz},\ref{jperpmod},\ref{sourcenot}) reveals that the only combinations of $\xi_7, ... ,\xi_{12}$ that need to be evaluated are $(\xi_8 - u^2 \xi_{12})$ and $\xi_9$.  These relevant combinations are given by
\begin{equation}
\int dk_z dk_T \frac{e^{i (z-u t) k_z}}{k_z^2 + \gamma^2 k_T^2}
\begin{bmatrix}
\gamma^2 J_1(\rho k_T) k_T^2\\
J_0(\rho k_T) k_z k_T
\end{bmatrix}
=
\begin{bmatrix}
\xi_9\\
\xi_8 - u^2 \xi_{12}
\end{bmatrix}
\end{equation}
where $\gamma^2 = (1 - u^2)^{-1}$.

The above expressions can be evaluated in a straightforward manner by first performing a contour integration in the $k_z$ variable.  The general form needed is
\begin{equation}
\int d k_z \frac{e^{\pm i k_z (z-u t)}}{(k_z^2 + k_T^2\gamma^2)}
\begin{bmatrix}
\gamma^2 k_T^2\\
k_z
\end{bmatrix}
= \pi e^{\mp(z-u t) k_T \gamma}
\begin{bmatrix}
\gamma \\
\pm i
\end{bmatrix}
\end{equation}
where $\pm$ refers to the sign of $(z - u t)$.  Similarly, one can use the relation
\begin{equation}
\begin{split}
\int &d k_T k_T e^{-k_T \gamma |z-ut|}
\begin{bmatrix}
J_1(\rho k_T)\\
J_0(\rho k_T)
\end{bmatrix} \\
&= \frac{1}{(\rho^2 + \gamma^2 (z-ut)^2)^{3/2}}
\begin{bmatrix}
\rho\\
\gamma |z-ut|
\end{bmatrix}
\end{split}
\end{equation}
to obtain the result
\begin{equation}\label{nascent}
\begin{bmatrix}
\xi_9\\
\xi_8 - u^2 \xi_{12}
\end{bmatrix}
=
\frac{\pi \gamma}{(\rho^2 + \gamma^2 (z-ut)^2)^{3/2}}
\begin{bmatrix}
\rho\\
i (z - u t)
\end{bmatrix}
\end{equation}

Before attempting an explicit evaluation of $\xi_1,...,\xi_{6}$ it is worthwhile to consider the general form of the source term given by (\ref{jz},\ref{jperpmod},\ref{sourcenot}).  It is interesting to note that (\ref{jz}) and (\ref{sourcenot}) have a similar form.  If one were to replace $ \xi_1 \rightarrow (u^2 \xi_2 - u^3 \xi_5 )$ in (\ref{jz}) then one would find the relation
\begin{equation}\label{global}
J_z \rightarrow u J_0.
\end{equation}
I am motivated to find a relationship between $\xi_1$, $\xi_2$, and $\xi_5$ that would allow me to write the source term in a form similar to that given by (\ref{global}).
I can (trivially) write
\begin{equation}
\begin{split}
\xi_1 &= \left(u^2 \xi_2 - u^3\xi_5\right) + \left(\xi_1 - u^2 \xi_2\right) + u^3\xi_5
\end{split}
\end{equation}
which allows
\begin{equation}\label{sourcezadv}
\begin{split}
J^{z} &= \frac{i g^2 {m^{2}_D}({Q_p^a})^2}{16 \pi^4}\left(u^2 \xi_9(u \xi_6  + \xi_4) - \left( \left(u^2 \xi_2 - u^3\xi_5\right) \right.\right.\\
&\left.\left. + \left(\xi_1 - u^2 \xi_2\right) + u^3\xi_5 \right)(\xi_8 - u^2 \xi_{12})\right)\\
&= u J^0 + 2(z - u t)\frac{d(\rho,z,t)}{\pi}\gamma\left(u^3\xi_5 + \left(\xi_1 - u^2 \xi_2\right)\right)
\end{split}
\end{equation}
and
\begin{equation}\label{sourcexadv}
\begin{split}
J^{x(y)} &= 2 x(y)\frac{d(\rho,z,t)}{\pi}\gamma\left(u^3 \xi_5  + ( \xi_1 - u^2 \xi_2)\right)
\end{split}
\end{equation}
where the function $d$ is given by
\begin{equation}
d(\rho,z,t) = \frac{\alpha_s ({Q_p^a})^2 m_{\rm D}^2}{8\pi(\rho^2 + \gamma^2(z-ut)^2)^{3/2}}.
\end{equation}
Notice that $J^{x(y)}$ now has the same form as $J^z - u J^0$ apart from the factors of $x$ and $(z - u t)$.  The source term can be written in the following compact form:
\begin{equation}\label{compact}
J^\nu(x) = \left(J^0(x),{\bf u} J^0(x) - {\bf J}_\text{v}\right)
\end{equation}
where
\begin{equation}\label{compactgrad}
\begin{split}
{\bf J}_\text{v} = \left(\bald{x} - \bald{u} t\right)\frac{2 \gamma}{\pi}d(\rho,z,t)\left(u^3 \xi_5  + ( \xi_1 - u^2 \xi_2)\right).
\end{split}
\end{equation}

\begin{figure*}
\centerline{
\includegraphics[width = 0.35\linewidth]{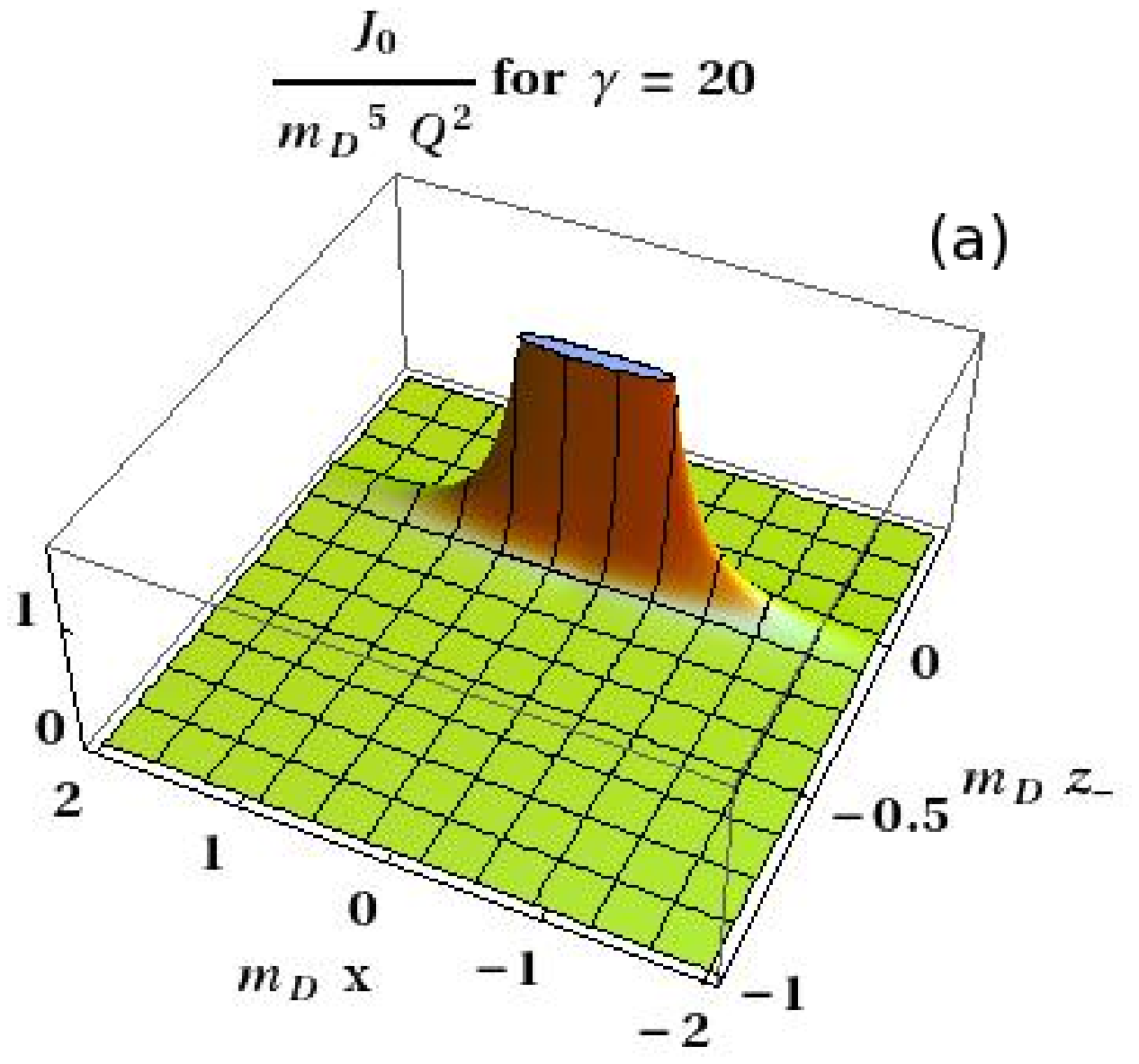} \hskip0.02\linewidth
\includegraphics[width = 0.35\linewidth]{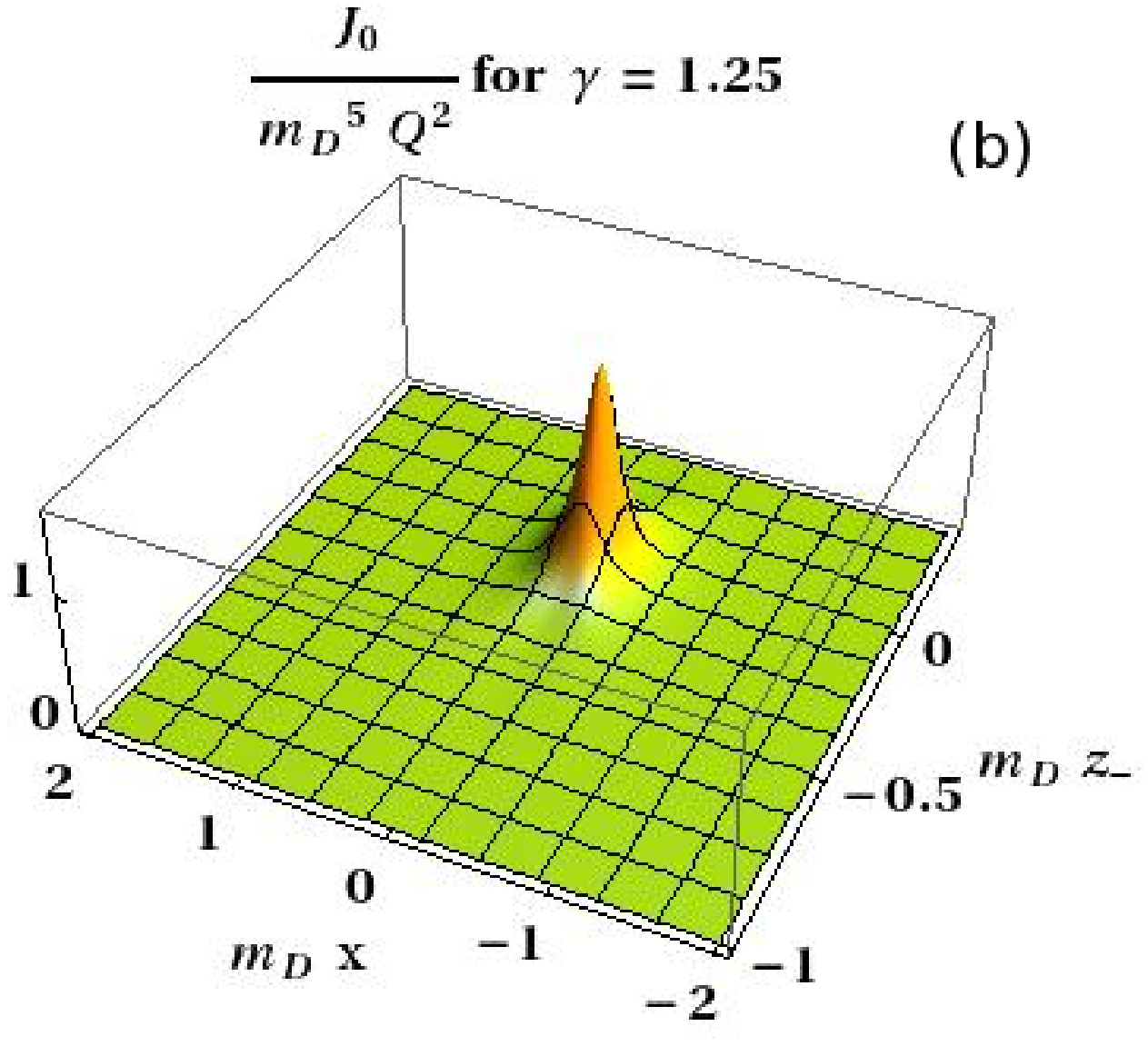}}
\caption{(color online) Three-dimensional plots of the scaled energy density deposited per unit time by a fast parton moving in the positive $z$ direction for $\alpha_s = 1/\pi$ and $N_c = 3$ [see (\ref{one})].  The plots are scaled by $m_{\rm D}^5 ({Q_p^a})^2  \approx 20.4 \text{ GeV}/\text{fm}^4$ for a gluon moving in a QGP at 200 MeV.  As one might expect, the distribution becomes Lorentz contracted for large values of $\gamma$.  (In the above plots, I have set $y = 0.4 \text{ GeV}^{-1}$ to avoid plotting the origin where the source is divergent.)}
\label{jnotbare}
\end{figure*}

I now continue by examining the terms $\xi_1,...,\xi_{6}$ which are considerably more difficult to evaluate than $\xi_7,...,\xi_{12}$ because of their dependence on the two $\Omega_i$ terms.  The result and method of their evaluation is discussed in detail in Appendix C.  Inserting the $\xi_i$ into (\ref{sourcenot}) and (\ref{compactgrad}) I eventually find for the source term
\begin{equation}\label{one}
\begin{split}
J^0(\rho,z,t) = &d(\rho,z,t)\gamma u^2 \\
&\times\left(1 - \frac{z_{-}}{(z_{-}^2 + \rho^2)}\left(z_{-} + \frac{\gamma u \rho^2}{\sqrt{\rho^2 + z_{-}^2 \gamma^2}}\right)\right)
\end{split}
\end{equation}
\begin{equation}\label{two}
\begin{split}
{\bf J}_\text{v}(\rho,z,t) = &\left(\bald{x} - \bald{u} t\right)\frac{\alpha_s ({Q_p^a})^2 m_{\rm D}^2}{8\pi \left(z_{-}^2+\rho^2\right)^2} \\
&\times\left(\frac{u^4 \rho^4+\left(z_{-}^2 \gamma ^2+\rho ^2\right) \left(2 z_{-}^2+\frac{\left(u^2+2\right)\rho^2}{\gamma^2}\right)}{\left(z_{-}^2 \gamma^2+\rho ^2\right)^2} \right.\\
&\left.-\frac{2 u z_{-}}{\gamma  \sqrt{z_{-}^2 \gamma^2+\rho ^2}}\right)
\end{split}
\end{equation}
where $z_{-} = (z - u t)$.

\begin{figure}
\centerline{
\includegraphics[width = 0.85\linewidth]{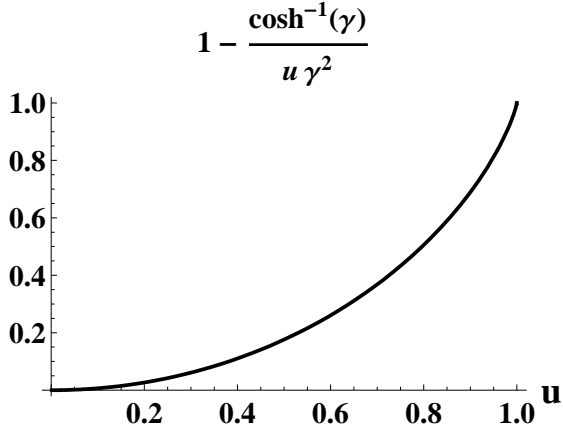}
}
\caption{Plot of the velocity dependence of $-d E/d x$ as given by (\ref{energyloss}) where $u$ is in units of $c =1$.}
\label{eloss}
\end{figure}

Equations (\ref{one}) and (\ref{two}), combined with (\ref{compact}), give the distribution of energy and momentum deposited into a perturbative QGP per unit time by a parton with constant velocity ${\bf u} = u \, \hat{\bf z}$ at position ${\bf x} = u t \, \hat{\bf z}$ in the absence of screening.  That this result can be expressed in relatively simple closed form expressions is remarkable.  The result for $J^0(\rho,z,t)$ is plotted in Fig. \ref{jnotbare} for two different values of $\gamma$.  The total energy deposited per unit time (or length) can be obtained by integrating (\ref{one}) over all space:
\begin{equation}\label{energyloss}
\begin{split}
-\frac{dE}{dx} &= \int d {\bf x}J^0(\rho,z,t)\\
&=  \frac{\alpha_s ({Q_p^a})^2 m_{\rm D}^2}{2} \ln \left[\frac{\rho_{\rm max}}{\rho_{\rm min}}\right]\left(1-\frac{y}{\gamma^2 u}\right)
\end{split}
\end{equation}
where $y = \cosh^{-1}(\gamma)$ is the rapidity and $\rho_{\rm max}$ and $\rho_{\rm min}$ are infrared and ultraviolet cut-offs for the $\rho$-integration.  This result matches the standard leading-logarithmic result \cite{Thoma:1991ea} for collisional energy loss in the relativistic ($u \rightarrow 1$) limit if one chooses $\rho_{\rm max} = 1/m_{\rm D}$ and $\rho_{\rm min} = 1/(2\sqrt{E_p T})$, where $E_p$ is the energy of the fast parton.  It is interesting to plot the $u$ dependence of (\ref{energyloss}) which is done in Fig. \ref{eloss}.  The total momentum deposited per unit time can likewise be obtained by integrating ${\bf J}(\rho,z,t)$ over all space (recall that ${\bf J} = {\bf u} J^0 - {\bf J}_\text{v}$).  One can verify explicitly that the source term satisfies the energy-momentum relation
\begin{equation}
\int d {\bf x}\, \bald{u}\cdot{\bf J}(\rho,z,t) = \int d{\bf x}\, J^0(\rho,z,t).
\end{equation}

\begin{figure}\centerline{
\includegraphics[width = 0.85\linewidth]{jnotscale.eps}
}
\caption{(color online) A plot of $J^0(\rho,z,t)/(\gamma u^2)$ as a function of $\gamma (z-u t)$ for the parameters $u = 0.99$, ${m_{\rm D}} = 0.4 \text{ GeV}$, $\alpha_s = 1/\pi$ and $\rho = 1$ GeV$^{-1}$.  If $J^0(\rho,z,t)$ depended only on the field configuration of the fast parton the curves should be identical.  As one can see, there is a slight shift in the curve when going from $\gamma = 1.25$ to $\gamma = 20$.  This shift reflects the dynamics of the medium in between the two interactions at $t'$ and $t$ (see discussion in the text).  The fact that the shift in the curve is small even though $\gamma$ has changed by a factor of 16 suggests the effect is small.}
\label{scaled}
\end{figure}

\begin{figure*}
\centerline{
\includegraphics[width = 0.35\linewidth]{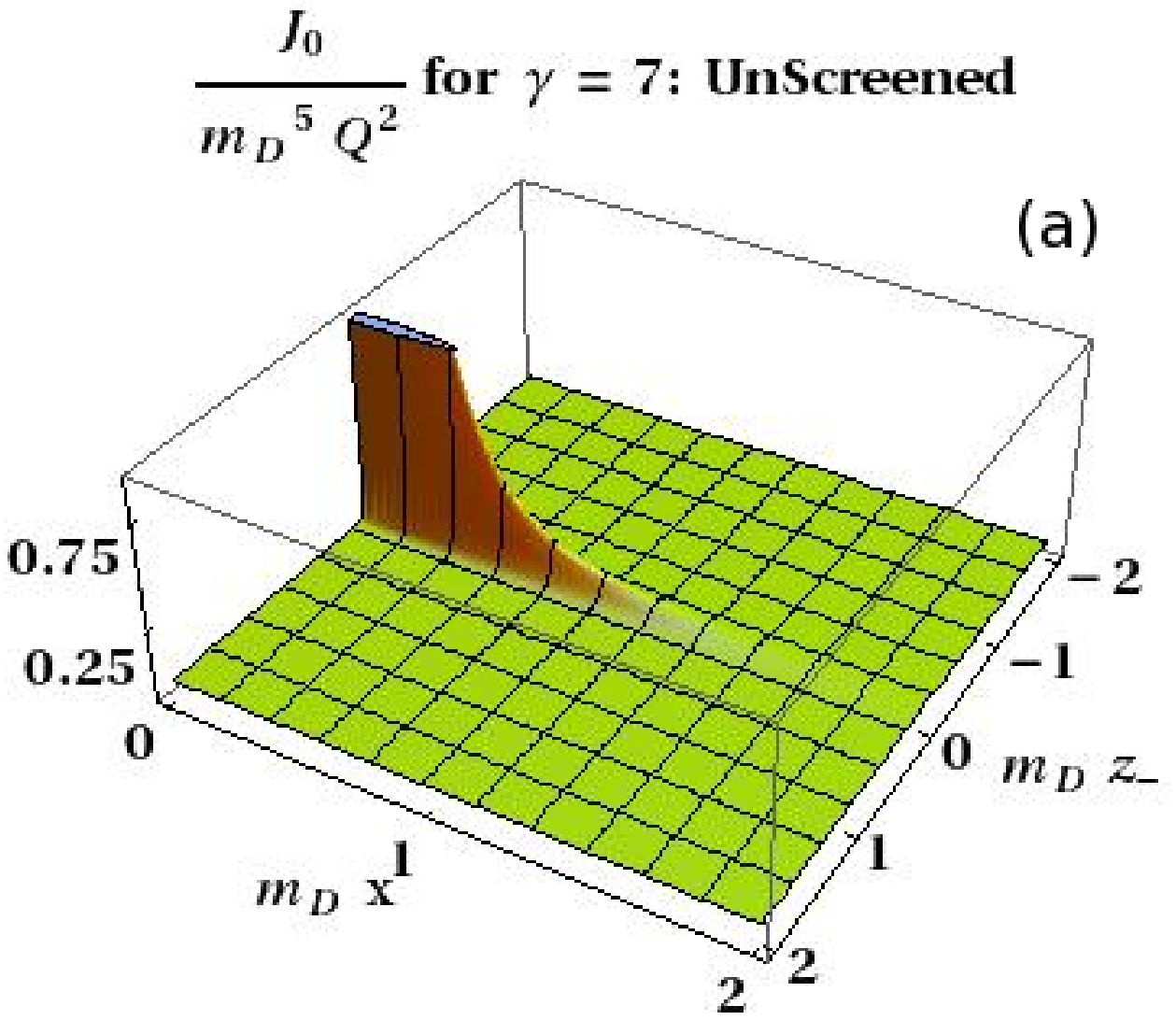} \hskip0.04\linewidth
\includegraphics[width = 0.35\linewidth]{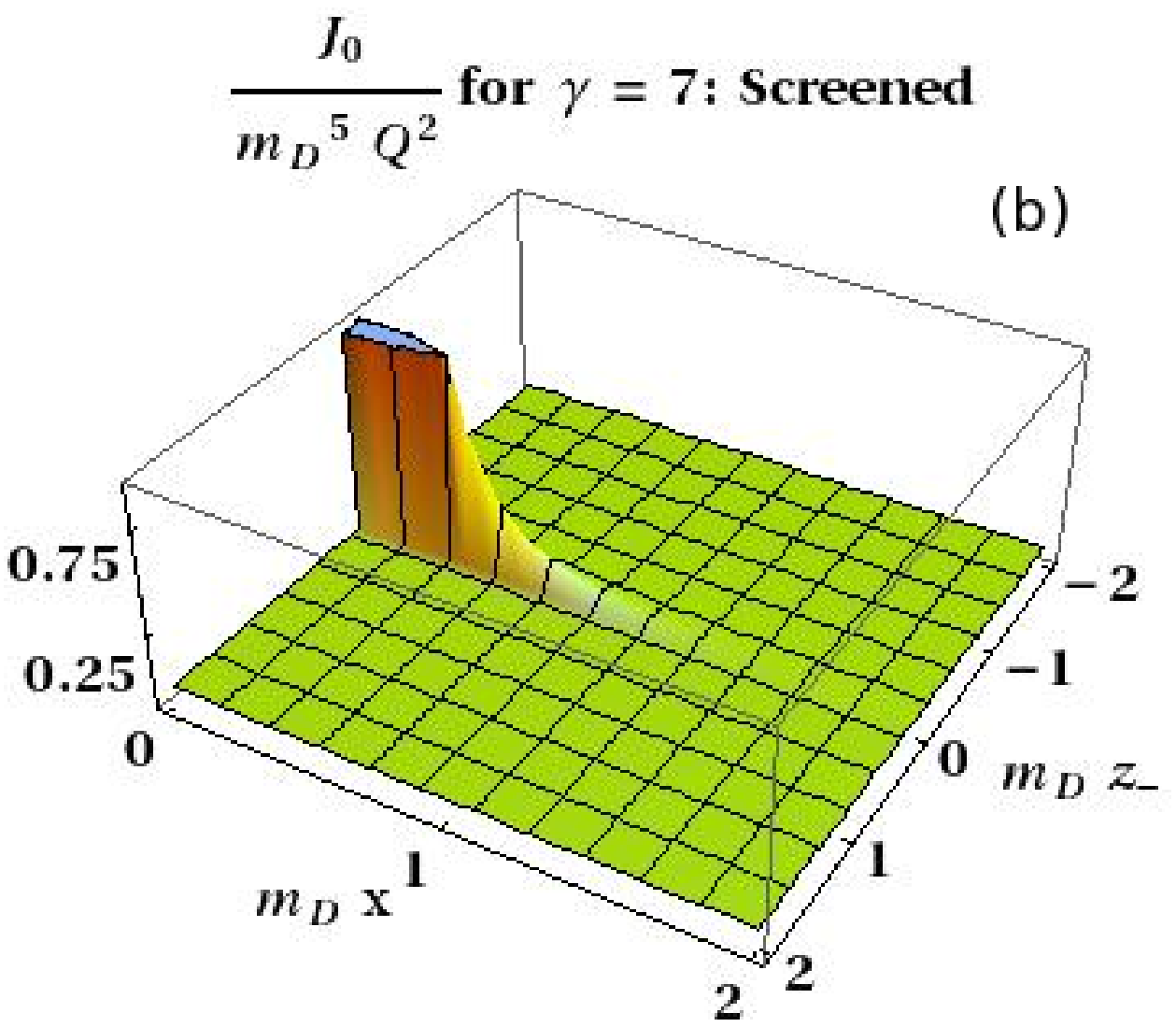}
}
\caption{(color online) Plots of the scaled energy density deposited per unit time by a parton moving in the positive $z$ direction with velocity $u = 0.99$ and $\alpha_s = 1/\pi$, both with and without medium screening (compare with Fig. \ref{jnotbare}).  There is a very similar structure in both cases; however, as one would expect, the screened result shows some damping in the transverse direction.}
\label{epsfigure}
\end{figure*}

\begin{figure*}
\centerline{
\includegraphics[width = 0.35\linewidth]{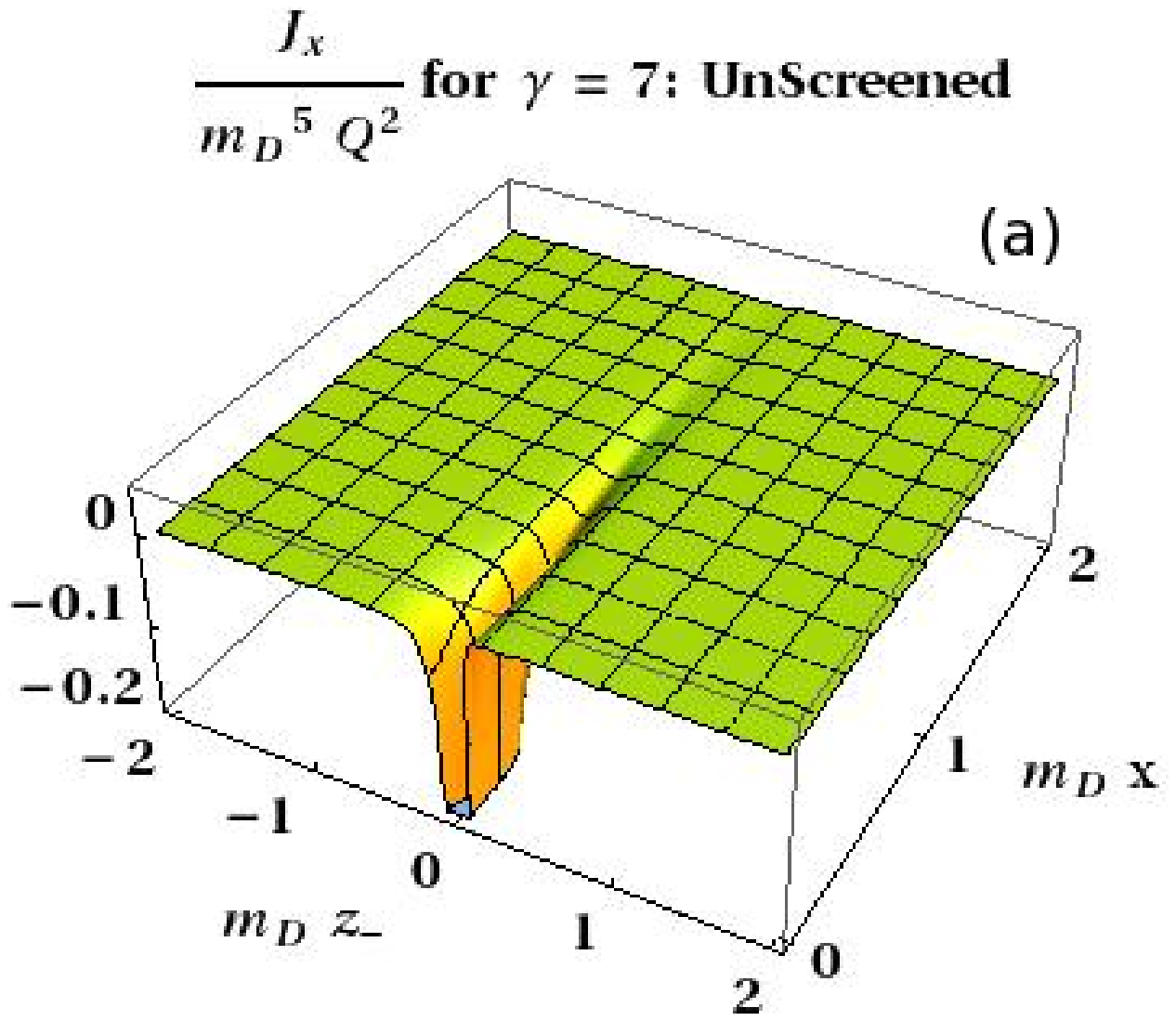} \hskip0.04\linewidth
\includegraphics[width = 0.35\linewidth]{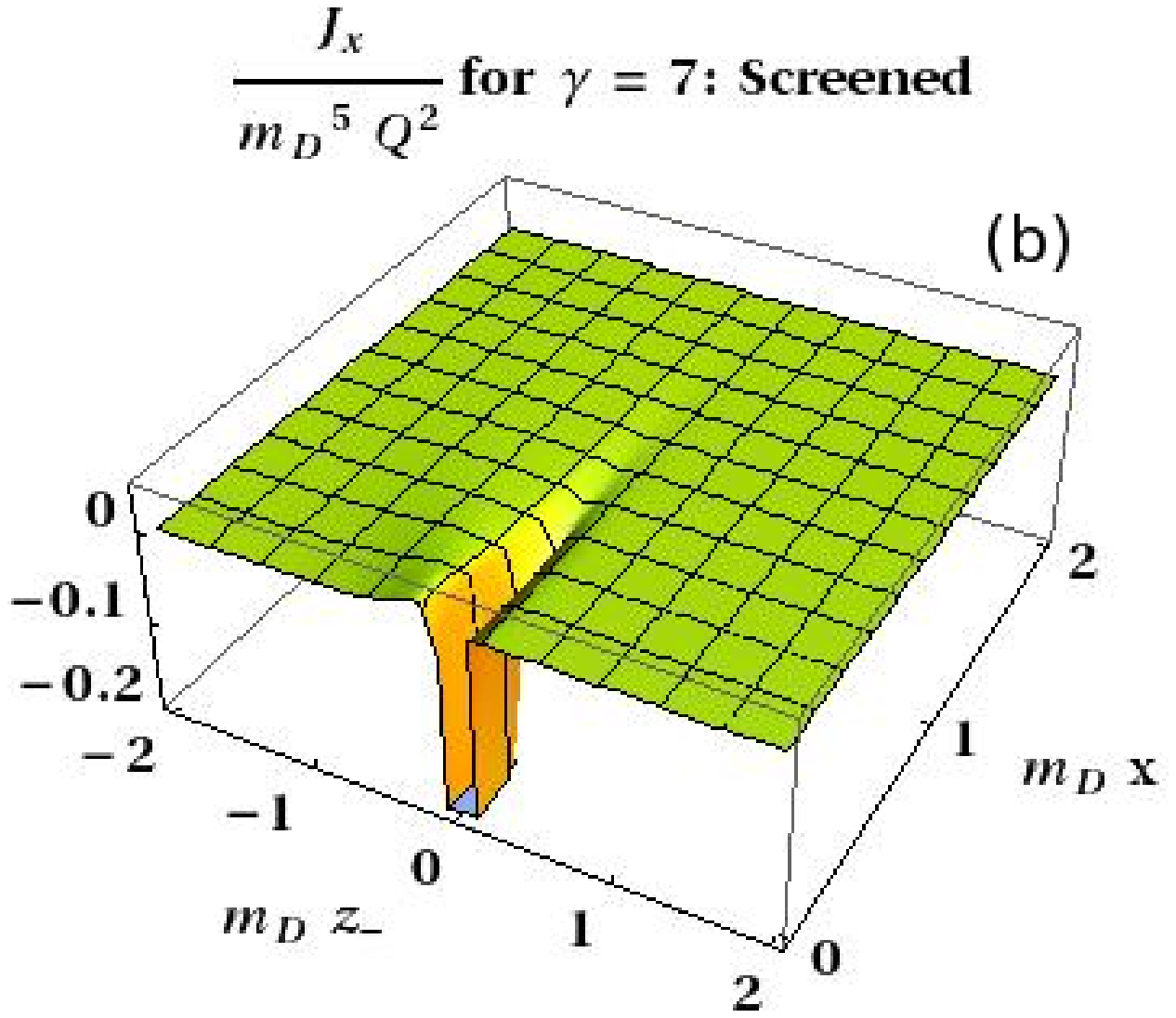}
}
\caption{(color online) Plots of the scaled transverse momentum density deposited per unit time both with and without medium screening included for the same parameters as in Fig. \ref{epsfigure}.  The structure shows noticeable differences in the two cases.  In particular, the screened result has a positive bump just behind the source parton which is absent in the unscreened result.}
\label{xfigure}
\end{figure*}

\begin{figure}
\centerline{
\includegraphics[width = 0.84\linewidth]{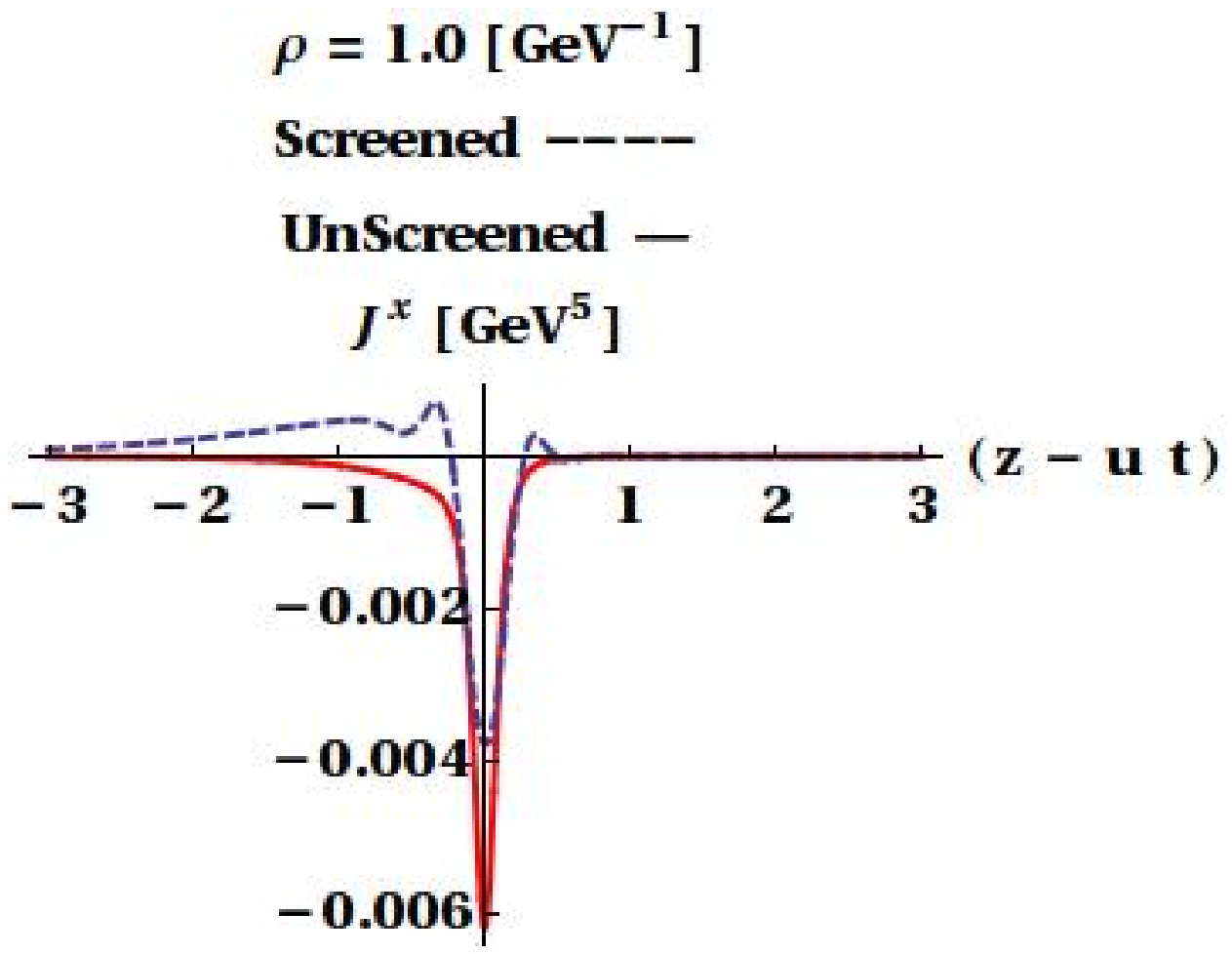}
}
\caption{(color online) Plot of transverse momentum density deposited per unit time by a gluon both with and without medium screening included (compare with Fig. \ref{xfigure}).  The plot is for $u = 0.99$, ${m_{\rm D}} = 0.4 \text{ GeV}$, and $\alpha_s = 1/\pi$.}
\label{linearjx}
\end{figure}

A careful observer may notice that the denominators of (\ref{one}) and (\ref{two}) contain terms involving both $\left(z_{-}^2 \gamma^2 +\rho^2\right)$, which encodes the Lorentz contraction of the fast parton's field configuration, and $\left(z_{-}^2 +\rho ^2\right)$, which is sensitive to the rest frame of the medium.  One may ask why the result, in the absence of dielectric screening, should depend on the rest frame of the medium.  The answer can be traced back to the $t'$ integration in (\ref{jnuadvanced}) which records the history of the fast parton's interactions with the medium.  In that expression the fast parton interacts with a medium particle at $t'$, which is expressed by $F_j(k')$, and again at $t$, which is expressed by $F_i(k)$.  In between $t'$ and $t$ the medium particle propagates with a momentum that depends on the properties of the medium [see (\ref{straight})].  Thus the second interaction, at $t$, depends on both the field configuration of the fast parton and the properties of the medium, even in the absence of screening.  One can observe this effect in Fig. \ref{scaled}, where $J^0(\rho,z,t)/(\gamma u^2)$ is plotted as a function of $\gamma z$ for fixed $\rho$.  If $J^0(\rho,z,t)$ depended only on the field configuration of the fast parton then the curves shown in Fig. \ref{scaled} should be identical for any velocity.  However, in the case of a highly relativistic source parton one sees a slight shift in the curve reflecting the dynamics of the medium in between the two interactions at $t'$ and $t$.

For a highly relativistic source parton Eqs. (\ref{one}) and (\ref{two}) can be written in a simplified form by considering that terms with $\left(z_{-}^2 \gamma^2+\rho ^2\right)$ in the denominator will be severely damped except for $z \approx u t$.  In this case, I keep $(z - u t)$ to first order (and drop terms proportional to $1/\gamma$, etc.) and the result becomes
\begin{eqnarray}
J^0(\rho,z,t) &=& d(\rho,z,t) \gamma u^2 \left( 1 - \frac{\gamma u \, z_-}{\sqrt{z_-^2 \gamma^2 + \rho^2}} \right)
\\
{\bf J}_\text{v}(\rho,z,t) &=& \left(\bald{x} - \bald{u} t \right) d(\rho,z,t) \frac{u^4}{\sqrt{z_-^2 \gamma^2 + \rho^2}}
\end{eqnarray}
In the next section, I consider the numerical evaluation of the source term with medium screening included and compare to the analytical result without screening.

\section{Result With Screening}\label{screen}

The evaluation of the source term (\ref{jperp},\ref{jz},\ref{jnot}) with medium screening included must be done numerically.  As mentioned before this amounts to evaluating the 12 $\xi_i$ terms listed in (\ref{kye}).  A discussion of the specific techniques used to perform the numerical integrations is given in Appendix D.

In this section all evaluations are for a parton moving along the positive $z$ axis with velocity $u = 0.99$ and I have set ${m_{\rm D}} = 0.4 \text{ GeV}$, $N_c = 3$ and $\alpha_s = 1/\pi$ (the choice of $u = 0.99$ corresponds to $\gamma\approx7$, or $E \approx 3$ GeV for the source gluon, which is likely too small a choice for the fast partons produced at RHIC but lends itself nicely to plotting the result).  The result for the energy density deposited per unit time both with and without medium screening is given in Fig. \ref{epsfigure}.  One sees a very similar structure in both cases except that the screened result is moderately damped in the transverse direction.  A similar plot is shown in Fig. \ref{xfigure} for the transverse momentum density deposited per unit time.  Here one finds more noticeable differences in the two cases.  For instance, the screened result has a positive bump just behind the source parton which is absent in the unscreened result.  This would correspond to matter being pushed outward, away from the source parton, in the vicinity of the bump.  The differences are further highlighted in Fig. \ref{linearjx}.  Plots of the parallel momentum density distribution, $J^z$, are not included because they are very similar to $J^0$ for a relativistic source.

\section{Discussion}

In this paper, I have derived the space-time distribution of energy and momentum deposited by a fast parton traversing a perturbative quark-gluon plasma by including physics at the Debye screening scale.  The approach has been to treat the fast parton as the source of an external color field perturbing the medium.  The final result depends on two parameters: the source strength $\alpha_s Q_p^2$ and the Debye mass $m_{\rm D} = \sqrt{N_c/3}\,g T$.

Several assumptions were made in deriving the result.  The first, and primary, of these is that the medium is perturbative in the coupling $\alpha_s$.  This assumption is reflected in the derivation of (\ref{mainvb}), the expansion of the path ordered gauge connection, $U_{ab}(x,x')$, and the choice of fields (\ref{efieldT} - \ref{bfield}).  Another assumption is that the medium is in local thermal equilibrium.  This is the basic assumption of hydrodynamics and is necessary to obtain (\ref{sourcehydro}).  I also assume that the typical momentum transfer between the source particle and the medium is small compared to the typical momentum of a medium particle.  This assumption allows the straight line approximation given by (\ref{straight}).  To get an idea of where this last assumption breaks down one can use the equipartition theorem for massless particles to find a relation between the typical momentum of a medium particle and the temperature: $\bar{p} = 3 T$.  For $T = 350$ MeV the typical momentum of a medium particle is about 1 GeV which places an upper limit on typical momentum transfers.  It follows that the straight line approximation breaks down somewhere on the order of $k \geq 1$ GeV$^{-1}$ (or distances less than 0.2 fm).

In the absence of the medium's response to screen the fields, it was possible to obtain an analytical result which is given by (\ref{one}) and (\ref{two}).  This result shows many similarities to the result obtained when the medium's response to screen the color field is included.  The most noticeable differences appear in the transverse momentum density distribution.  The total energy deposited per unit time is obtained by integrating $J^0(\rho,z,t)$ over all space and is given by (\ref{energyloss}) for the unscreened result.

The reader interested in reproducing the numerical results for the screened source term should consult Appendix D, where the specific techniques used to perform the numerical integrations are discussed.

{\it Acknowledgments:} I thank Berndt M\"uller for many discussions and advice.  I also thank Stanislaw Mrowczynski for helpful comments, and Jorge Noronha for pointing our an error in an earlier version of the manuscript.  This work was supported in part by the U.~S.~Department of Energy under Grant DE-FG02-05ER41367.

\section*{Appendix A: Derivation of (\ref{efieldT}) - (\ref{bfield})}  
\renewcommand{\theequation}{A-\arabic{equation}}
  \setcounter{equation}{0}  
\renewcommand{\thefigure}{A-\arabic{figure}}
  \setcounter{figure}{0}  

In the HTL approximation, the source fields obey Maxwell's equations, which are given by
\begin{eqnarray}\label{maxwell}
{\bm\nabla}\cdot\bald{D}(\bald{x},t) &=& \rho(\bald{x},t) \\
{\bm\nabla}\times\bald{E}(\bald{x},t) &=& -\frac{\partial\bald{B}(\bald{x},t)}{\partial t} \\
{\bm\nabla}\cdot\bald{B}(\bald{x},t) &=& 0 \\
\label{maxwell1}
{\bm\nabla}\times\bald{B}(\bald{x},t) &=& \bald{j}(\bald{x},t) + \frac{\partial\bald{D}(\bald{x},t)}{\partial t}
\end{eqnarray}
where the source density is
\begin{eqnarray}
\label{rho}
\rho(\bald{x},t) &=& g Q_p \delta(\bald{x} - \bald{u} t) \\
\label{jay}
\bald{j}({\bf x},t) &=& \bald{u}\rho(\bald{x},t)
\end{eqnarray}
(in the above equations $\bald{u}$ is the source particle's velocity and I have suppressed the color index, which I will restore in the final result).  It is convenient to solve for the fields by taking the Fourier transform of Maxwell's equations.  Using the general rule
\begin{equation}
F({\bf x},t) = \frac{1}{(2 \pi)^4}\int d^3 k \int d \omega \,e^{i \bald{k}\cdot\bald{x} - i \omega t} F({\bf k},\omega)
\end{equation}
it is found that
\begin{eqnarray}\label{maxwellfour}
i\bald{k}\cdot\bald{D}(\bald{k},\omega) &=& \rho(\bald{k},\omega) \\
i\bald{k}\times\bald{E}(\bald{k},\omega) &=& i\omega\bald{B}(\bald{k},\omega) \\
i\bald{k}\cdot\bald{B}(\bald{k},\omega) &=& 0 \\
\label{maxwellfour1}
i\bald{k}\times\bald{B}(\bald{k},\omega) &=& \bald{j}(\bald{k},\omega) -i\omega\bald{D}(\bald{k},\omega)
\end{eqnarray}
where
\begin{eqnarray}
\label{rhofour}
\rho(\bald{k},\omega) &=& 2 \pi g Q_p \delta(\omega - \bald{u}\cdot\bald{k}) \\
\label{jayfour}
\bald{j}({\bf k},\omega) &=& \bald{u}\rho(\bald{k},\omega).
\end{eqnarray}
The function $\bald{D}$ is related to $\bald{E}$ through the dielectric tensor $\epsilon_{i j}$:
\begin{equation}
D_i = \epsilon_{ij}E_j
\end{equation}
(the summation over $j$ is implied and I now suppress the $(\bald{k},\omega)$ notation).  In an isotropic medium the dielectric tensor can be decomposed into transverse and longitudinal parts such that
\begin{equation}
\epsilon_{ij} = \epsilon_L\frac{k_i k_j}{k^2} + \epsilon_T\left(\delta_{ij} - \frac{k_i k_j}{k^2}\right)
\end{equation}
and $\bald{D}$ is now written as
\begin{equation}
\bald{D} = \epsilon_L\bald{E}_L + \epsilon_T\bald{E}_T
\end{equation}
where $ {\bf E}_L = \hat{{\bf k}}(\hat{{\bf k}}\cdot {\bf E})$ and $ {\bf E}_T = {\bf E} - {\bf E}_L$.  One should note that $\bald{k}\cdot\bald{E}_T = 0$ and $\bald{k}\times\bald{E}_L = 0$. Maxwell's equations now take the form
\begin{eqnarray}
\label{maxone}
i\epsilon_L\bald{k}\cdot\bald{E}_L &=& \rho \\
\label{maxtwo}
i\bald{k}\times\bald{E}_T &=& i\omega\bald{B} \\
\label{maxthree}
i\bald{k}\cdot\bald{B} &=& 0 \\
\label{maxfour}
i\bald{k}\times\bald{B} &=& \bald{u}\rho -i\omega(\epsilon_L\bald{E}_L + \epsilon_T\bald{E}_T).
\end{eqnarray}
An equation for $\bald{B}$ can be obtained by crossing $\bald{k}$ into (\ref{maxfour}) which combines with (\ref{maxtwo}) and (\ref{maxthree}) to yield (with color index restored)
\begin{equation}
\label{bmax}
\bald{B}^a = \frac{2 \pi i g Q^a_p (\bald{k}\times\bald{u})}{k^2 - \epsilon_T\omega^2}\delta(\omega - \bald{u}\cdot\bald{k}).
\end{equation}
One can then obtain $\bald{E}_T$ by crossing $\bald{k}$ into (\ref{maxtwo})
\begin{equation}
\label{eTmax}
\bald{E}^a_T = \frac{\omega}{k^2}\bald{B}\times\bald{k} = \frac{2 \pi i g Q^a_p \omega\left(\bald{u}k^2 - \bald{k}(\bald{k}\cdot\bald{u})\right)}{k^2(k^2 - \epsilon_T\omega^2)}\delta(\omega - \bald{u}\cdot\bald{k}).
\end{equation}
Combining (\ref{bmax}) and (\ref{eTmax}) with (\ref{maxfour}) yields
\begin{equation}\label{eLmax}
\begin{split}
\bald{E}^a_L &= (-1)\frac{i}{\epsilon_L\omega}\left(\bald{u}\rho + i\frac{k^2}{\omega}\bald{E}_T -i\omega \epsilon_T\bald{E}_T)\right) \\
&= -\frac{2 \pi i g Q^a_p (\bald{k}\cdot\bald{u})}{\epsilon_L \omega k^2}\bald{k}\delta(\omega - \bald{u}\cdot\bald{k}).
\end{split}
\end{equation}
These results can be back transformed to position space to yield (\ref{efieldT} - \ref{bfield}).

\section*{Appendix B: Explicit evaluation of the $\hat{\bf v}$ integration}  
\renewcommand{\theequation}{B-\arabic{equation}}
  \setcounter{equation}{0}  
\renewcommand{\thefigure}{B-\arabic{figure}}
  \setcounter{figure}{0}  

In going from Eq. (\ref{jnuadv}) to (\ref{jnuone}) and (\ref{jnutwo}) I performed an integration over $\hat{ \bf v}$ by choosing a frame in which ${\bf k'} = k'\hat{ \bf z}$, performing the integral, and then rotating back into a frame in which ${\bf k'}$ is arbitrary.  In this appendix the explicit details involved will be given.  The relevant starting point is
\begin{equation}
\begin{split}\label{vhat}
\int d \hat{{\bf v}}\frac{\hat{{\bf v}}\cdot {\bf E'}^a\left(\delta_{0\nu}\hat{{\bf v}}\cdot{\bf E}^{a} + \delta_{i\nu} \left({E^{a}_i} + ({\hat{{\bf v}}\times{\bf B}})^{a}_i\right)\right)}{4 \pi(\omega'-{\bf k'}\cdot\hat{ \bf v} + i\epsilon)}
\end{split}
\end{equation}
where ${\bf E'}^a$ is short for ${\bf E}^a(k')$, etc.  There are two basic integrals to consider, which have the form
\begin{eqnarray}
\label{first} \int d \hat{ \bf v} \frac{ \hat{ \bf v}_j}{4 \pi(\omega' - {\bf k'}\cdot\hat{ \bf v} + i \epsilon)} \equiv I_j\\  \label{second} \int d \hat{ \bf v} \frac{ \hat{ \bf v}_j\hat{ \bf v}_m}{4 \pi(\omega' - {\bf k'}\cdot\hat{ \bf v} + i \epsilon)} \equiv I_{jm}.
\end{eqnarray}
Now specify that ${\bf k'} = k'\hat{ \bf z}$ and consider that since $\hat{ \bf v} = (\sin\theta\cos\phi,\sin\theta\sin\phi,\cos\theta)$ the integral over $d\phi$ will ensure that only $I_z$ survives in (\ref{first}).  Making the replacement $d \hat{\bf v} \rightarrow \int d \theta d\phi \sin \theta$ gives
\begin{equation}\label{ix}
\begin{split}
I_z &= \int d \theta d\phi \frac{ \sin\theta\cos\theta}{4 \pi(\omega' - k'\cos\theta + i \epsilon)} \\
&= \frac{1}{2}\int_{-1}^1\frac{d \xi\text{ }\xi}{(\omega' - k'\xi + i \epsilon)} \equiv \Omega_1.
\end{split}
\end{equation}
Similarly, in (\ref{second}) the integral over $d\phi$ ensures only $I_{xx}$, $I_{yy}$, and $I_{zz}$ contribute:
\begin{equation}\label{ixx}
I_{xx} = I_{yy} = \frac{1}{4}\int_{-1}^1\frac{d \xi(1 - \xi^2)}{(\omega' - k'\xi + i \epsilon)}\equiv \Omega_2
\end{equation}
and
\begin{equation}\label{izz}
I_{zz} = \frac{1}{2}\int_{-1}^1\frac{d \xi\text{ }\xi^2}{(\omega' - k'\xi + i \epsilon)}\equiv \Omega_3.
\end{equation}
The $\Omega_i$ terms are evaluated with the result
\begin{equation}
\Omega_1(k) = \frac{1}{2k}\left(\frac{\omega}{k}\ln{\left[\frac{k + \omega}{k - \omega}\right]}-\frac{\pi i \omega}{k}-2\right)
\end{equation}
\begin{equation}
\begin{split}
\Omega_2 (k) = &\frac{1}{4 k}\left((1-\frac{\omega^2}{k^2})\ln{\left[\frac{k + \omega}{k - \omega}\right]}\right. \\
&\left.-\pi i (1-\frac{\omega^2}{k^2}) + 2\frac{\omega}{k}\right)
\end{split}
\end{equation}
\begin{equation}
\Omega_3(k) = \frac{\omega}{2 k^2}\left(\frac{\omega}{k}\ln{\left[\frac{k + \omega}{k - \omega}\right]}-\frac{\pi i \omega}{k}-2\right)=\frac{\omega}{k}\Omega_1(k).
\end{equation}
Equations (\ref{ix} - \ref{izz}) can be rewritten as
\begin{equation}\label{ixi}
I_j = \delta_{j z}\Omega_1
\end{equation}
\begin{equation}\label{ixxi}
\begin{split}
I_{jm} &= \left(\delta_{j x} \delta_{m x} + \delta_{j x} \delta_{m x}\right)\Omega_2 + \delta_{j z}\delta_{m z}\Omega_3 \\
&= \delta_{jm}\Omega_2 + \delta_{j z}\delta_{m z}(\frac{\omega}{k}\Omega_1 - \Omega_2).
\end{split}
\end{equation}

In the above expressions the subscript $z$ denotes the orientation of the ${\bf k'}$ vector.  Generalizing to an arbitrary ${\bf k'}$ is done by making the replacement $\delta_{j z}\rightarrow{ \bf k'}_j$ in (\ref{ixi}) and (\ref{ixxi}).  Going back to (\ref{vhat}) (and suppressing the color index) I find
\begin{equation}
\begin{split}
\int d \hat{{\bf v}}&\frac{\hat{{\bf v}}\cdot {\bf E'}^a\left(\delta_{0\nu}\hat{{\bf v}}\cdot{\bf E}^{a} + \delta_{i\nu} \left({E^{a}_i} + ({\hat{{\bf v}}\times{\bf B}})^{a}_i\right)\right)}{4 \pi(\omega'-{\bf k'}\cdot\hat{ \bf v} + i\epsilon)} = \\ \delta_{0\nu}&{E'_j}{E_m}I_{jm} + \delta_{i\nu} \left({E'_j}{E_i}I_{j} + \epsilon_{imk}I_{j m}{E'_j}{B_k}\right) = \\
\delta_{0\nu}&\left(({\bf E'}\cdot{\bf E})\Omega_2 + (\hat{ \bf k'}\cdot{\bf E'})(\hat{ \bf k'}\cdot{\bf E})(\frac{\omega}{k}\Omega_1 - \Omega_2)\right) + \\
\delta_{i\nu} &\left((\hat{ \bf k'}\cdot{\bf E'}){E_i}\Omega_1 + \right. \\
&\left.\left(({\bf E'}\times{\bf B})_i\Omega_2 + (\hat{ \bf k'}\cdot{\bf E'})(\hat{\bf k'}\times{\bf B})_i(\frac{\omega}{k}\Omega_1 - \Omega_2)\right)\right)
\end{split}
\end{equation}
which is the result given in (\ref{jnuone}) and (\ref{jnutwo}).

\section*{Appendix C: Evaluation of $\xi_1,...,\xi_6$}  
\renewcommand{\theequation}{C-\arabic{equation}}
  \setcounter{equation}{0}  
\renewcommand{\thefigure}{C-\arabic{figure}}
  \setcounter{figure}{0}  
In Sec. \ref{unscreen}, I put off the explicit evaluation of the terms $\xi_1,...,\xi_{6}$ and directly went to the results (\ref{one}) and (\ref{two}).  In this appendix, I will give a detailed derivation of the expressions:
\begin{equation}\label{hard}
\begin{split}
\begin{bmatrix}
\xi_1\\
\xi_2\\
\xi_3\\
\xi_4
\end{bmatrix}
&\equiv
\int dk_z dk_T \frac{e^{i (z-u t) k_z} \Omega_1(k) }{k D_L(k)}
\begin{bmatrix}
J_0(\rho k_T)k_T k^2 \\
J_0(\rho k_T) k_T k_z^2\\
J_0(\rho k_T) k_T^3\\
J_1(\rho k_T) k_z k_T^2
\end{bmatrix}
\\
\begin{bmatrix}
\xi_5\\
\xi_6
\end{bmatrix}
&\equiv
\int dk_z dk_T \frac{e^{i (z-u t) k_z} \Omega_2(k) }{k^2 D_T(k)}
\begin{bmatrix}
J_0(\rho k_T)k_z k_T^3\\
J_1(\rho k_T) k_T^2 k_z^2
\end{bmatrix}.
\end{split}
\end{equation}
Here, I am interested in the unscreened source term which is found by setting $D_L(k) = k^2$ and $D_T(k) = k^2 - u^2 k_z^2$.
I start by changing to polar coordinates (i.e., $k$ and $\theta$, where $k_z = k \cos\theta$ and $k_T = k \sin\theta$) in which case the two $\Omega$ terms can be written as
\begin{eqnarray}
\Omega_1(k) = \frac{G_1(\theta)}{2 k} \\ \Omega_2(k) = \frac{G_2(\theta)}{4 k}
\end{eqnarray}
where
\begin{equation}
G_1(\theta) = 2 u\cos\theta\tanh^{-1}[u\cos\theta] - i \pi u \cos\theta -2
\end{equation}
and
\begin{equation}
\begin{split}
G_2(\theta) = 2 \tanh^{-1}&[u\cos\theta](1 - u^2\cos^2\theta) \\
&- i \pi (1 - u^2\cos^2\theta) + 2 u\cos\theta.
\end{split}
\end{equation}
Defining $\Lambda(\theta)$ as some generic function, I now write the general form of the $k$ integration as
\begin{equation}
\xi_i = \int d k d\theta \sin\theta e^{i k((z-u t) \cos\theta)}
\begin{bmatrix}
J_0(\rho k\sin\theta)\\
J_1(\rho k\sin\theta)
\end{bmatrix}\Lambda_i(\theta)
\end{equation}
which can be evaluated using the relations \cite{magnus}:
\begin{eqnarray}
\int_0^\infty d k \,e^{i k y}J_0(k b) = \frac{1}{\sqrt{b^2 - (y + i\epsilon)^2}} \\
\int_0^\infty d k \,e^{i k y}J_1(k b) = \frac{1}{b}\left(1 + \frac{i y}{\sqrt{b^2 - (y + i\epsilon)^2}}\right)
\end{eqnarray}
where $b > 0$ and $\epsilon$ is a positive infinitesimal quantity.  Defining $a \equiv (z - u t)/\rho$ and using the above integrals allows
\begin{equation}
\begin{split}
\xi_i &= \int d k d\theta \sin\theta e^{i k((z-u t) \cos\theta)}
\begin{bmatrix}
J_0(\rho k\sin\theta)\\
J_1(\rho k\sin\theta)
\end{bmatrix}\Lambda_i(\theta) \\
&=
\int_0^\pi d\theta \frac{1}{\rho}
\begin{bmatrix}
\frac{\sin\theta}{\sqrt{\sin^2\theta-(a \cos\theta + i\epsilon)^2}}\\
1 + \frac{i z \cos\theta}{\rho\sqrt{\sin^2\theta - (a\cos\theta + i\epsilon)^2}}
\end{bmatrix}\Lambda_i(\theta)\\
&\equiv \int_0^\pi d\theta \frac{1}{\rho}
\begin{bmatrix}
U(\theta)\\
L(\theta)
\end{bmatrix}\Lambda_i(\theta).
\end{split}
\end{equation}
where I have set $t = 0$ for conciseness.  Explicitly, the terms look like
\begin{equation}
\begin{split}
\begin{bmatrix}
\xi_1\\
\xi_2\\
\xi_4
\end{bmatrix}
&=
\int_0^\pi d\theta \frac{1}{2 \rho}G_1(\theta)
\begin{bmatrix}
U(\theta)\\
U(\theta)\cos^2\theta\\
L(\theta)\cos\theta\sin\theta
\end{bmatrix}
\\
\begin{bmatrix}
\xi_5\\
\xi_6
\end{bmatrix}
&=
\int_0^\pi d\theta \frac{1}{4 \rho}G_2(\theta)
\begin{bmatrix}
U(\theta)\cos\theta\sin^2\theta\\
L(\theta)\cos^2\theta\sin\theta
\end{bmatrix}.
\end{split}
\end{equation}
where $\xi_3$ is omitted since it will not appear in the final expressions.

It is imperative to consider symmetry before trying to evaluate the various terms.  For instance, in $\xi_1,...,\xi_4$ the imaginary part of $U(\theta)$ and the real part of $L(\theta)$ only need to multiply against the term $- i \pi u \cos\theta$ due to symmetry considerations.  Similarly, the real part of $U(\theta)$ and the imaginary part of $L(\theta)$ do not need to multiply against $- i \pi u \cos\theta$.  It turns out that due to symmetry the terms involving the imaginary part of $U(\theta)$ and the real part of $L(\theta)$ are easier to evaluate than those involving the real part of $U(\theta)$ and the imaginary part of $L(\theta)$.  It makes sense to first get these simpler expressions out of the way and then focus on the more tedious ones.  I introduce the notation $\xi_i = {\xi_i}_a + {\xi_i}_b$, where ${\xi_i}_a$ now denotes terms involving the imaginary part of $U(\theta)$ and the real part of $L(\theta)$.  Using {\it Mathematica 6.0} these terms can be evaluated and I find
\begin{equation}
\begin{split}
\begin{bmatrix}
{\xi_1}_a\\
{\xi_2}_a\\
{\xi_4}_a
\end{bmatrix}
&=
\begin{bmatrix}
\frac{\pi a u}{\rho\chi^2}\\
\frac{\pi a u(3 + a^2)}{3 \rho\chi^4}\\
- \frac{i \pi u}{3\rho}\left(1 - \frac{z}{\rho}\frac{a(3 + a^2)}{\chi^4}\right)
\end{bmatrix}
\\
\begin{bmatrix}
{\xi_5}_a\\
{\xi_6}_a
\end{bmatrix}
&=
\begin{bmatrix}
\frac{\pi a^3}{3 \rho \chi^4}\\
- \frac{i \pi }{6\rho}\left(1 - \frac{z}{\rho}\frac{a(3 + a^2)}{\chi^4}\right)
\end{bmatrix}
\end{split}
\end{equation}
where $\chi\equiv\sqrt{1 + a^2}$.

Now I turn to the more involved ${\xi_i}_b$ terms which involve the real part of $U(\theta)$ and the imaginary part of $L(\theta)$. It is convenient to note that
\begin{eqnarray}
\text{Re}[U(\theta)] = \sin\theta\,\text{Re}\left[\frac{1}{\sqrt{\sin^2\theta - a^2\cos^2\theta}}\right]\\
\text{Im}[L(\theta)] = \frac{z\cos\theta}{\rho}\text{Re}\left[\frac{1}{\sqrt{\sin^2\theta - a^2\cos^2\theta}}\right].
\end{eqnarray}
For the sake of bookkeeping, I note there are 7 distinct integrations which need to be performed.  These terms are given by the real parts of
\begin{equation}
\varphi_i = \int d\theta\frac{\sin\theta}{\sqrt{\sin^2\theta - a^2\cos^2\theta}}T_i
\end{equation}
where
\begin{equation}
\begin{split}
\begin{bmatrix}
T_1\\
T_2\\
T_3\\
T_4\\
T_5\\
T_6\\
T_{7}
\end{bmatrix}
&\equiv
\begin{bmatrix}
1\\
\cos^2\theta\\
\cos\theta\tanh^{-1}[u\cos\theta]\\
\cos^3\theta\tanh^{-1}[u\cos\theta]\\
\cos\theta\sin^2\theta\tanh^{-1}[u\cos\theta]\\
\cos^2\theta\sin^2\theta/(1 - u^2\cos^2\theta)\\
\cos^4\theta/(1 - u^2\cos^2\theta)
\end{bmatrix}
\end{split}.
\end{equation}
In terms of these the ${\xi_i}_b$ are
\begin{equation}
\begin{split}
\begin{bmatrix}
{\xi_1}_b\\
{\xi_2}_b\\
{\xi_4}_b\\
{\xi_5}_b\\
{\xi_6}_b
\end{bmatrix}
&=
\frac{1}{\rho}
\begin{bmatrix}
u \varphi_3 - \varphi_1\\
u \varphi_4 - \varphi_2\\
\frac{z}{\rho}(u i \varphi_4 - i \varphi_2)\\
(\varphi_5 + u \varphi_6)/2\\
\frac{z}{2\rho}(i  \varphi_4 + i u \varphi_{7})
\end{bmatrix}
\end{split}.
\end{equation}
Using {\it Mathematica 6.0} I find for the real parts of the $\varphi$ terms:
\begin{equation}
\varphi_1 = \frac{\pi}{\chi}
\end{equation}
\begin{equation}
\varphi_2 = \frac{\pi}{2 \chi^3}
\end{equation}
\begin{equation}
\begin{split}
\varphi_3 = \frac{\pi}{u \chi^2}\left(\chi - \sqrt{a^2 + 1/\gamma^2}\right)
\end{split}
\end{equation}
\begin{equation}
\begin{split}
\varphi_4 = \frac{(\chi^2 + 2 u^2)}{3 u^2\chi^2}\varphi_3 - \frac{\pi}{6 u \chi^3}
\end{split}
\end{equation}
\begin{equation}
\begin{split}
\varphi_5 = \varphi_3 - \varphi_4
\end{split}
\end{equation}
\begin{equation}
\begin{split}
\varphi_6 = \frac{\pi}{\gamma^2 u^4}\left(\frac{2 \chi^2 + \gamma^2 u^2}{2 \chi^3} - \frac{1}{\sqrt{a^2 + 1/\gamma^2}}\right)
\end{split}
\end{equation}
\begin{equation}
\begin{split}
\varphi_{7} = \frac{\pi}{u^2}\left(-\frac{1}{\chi} + \frac{1}{\sqrt{a^2 + 1/\gamma^2}} \right) -\varphi_6
\end{split}
\end{equation}
Combining the above results I find
\begin{widetext}
\begin{equation}\label{angie}
\begin{split}
&\xi_1 = \frac{-\pi \left(\sqrt{a^2 + 1/\gamma^2} - a u\right)}{\rho \chi^2}\\
&\xi_5 = \frac{\pi  \left(2 \sqrt{a^2+\frac{1}{\gamma^2}} \gamma^2 \left(a^3 u^3 + \chi^3\right) + \left(1 + a^2 \gamma^2\right)\left(2 u^2 + \chi^2(1 - 3 u^2)\right) - 3 \chi^4\right)}{6 \rho u^3 \chi^4 \sqrt{a^2+\frac{1}{\gamma^2}} \gamma^2} \\
&\xi_1 - u^2 \xi_2 = -\frac{\pi}{3 \rho \chi^4}\left(a u^3\left(2 + \chi^2\right) + \chi^2\left(\chi - 3 a u\right) + 2 \sqrt{a^2+\frac{1}{\gamma ^2}}(\chi^2 - u^2)\right) \\
&u \xi_2 - u^2\xi_5 = \frac{\pi  u \left(a \left(-2 a \gamma^2 + 2 u \gamma^2\sqrt{a^2+\frac{1}{\gamma ^2}}  + a\right)-1\right)}{2 \chi^4 \sqrt{a^2+\frac{1}{\gamma ^2}} \gamma^2} \\
&u\xi_6 + 2 \xi_4 = \frac{i \pi  z \left(-a^2 + \left(a^2+3\right) u a \sqrt{a^2+\frac{1}{\gamma ^2}} + 2 u^2-1\right)}{2 \chi^4 \sqrt{a^2+\frac{1}{\gamma ^2}} \rho }-\frac{i \pi  u}{2}
\end{split}
\end{equation}
\end{widetext}
which are combined with (\ref{nascent}) to yield (\ref{one}) and (\ref{two}).

\section*{Appendix D: Numerically Integrating Equation (\ref{kye})}  
\renewcommand{\theequation}{D-\arabic{equation}}
  \setcounter{equation}{0}  
\renewcommand{\thefigure}{D-\arabic{figure}}
  \setcounter{figure}{0}  

As mentioned in Sec.\ref{screen}, the evaluation of the 12 $\xi_i$ terms listed in (\ref{kye}) with medium screening included must be done numerically.  In this appendix I will discuss the specific techniques used to perform these numerical integrations.  The $\xi_i$ terms are given by
\begin{equation}
\begin{bmatrix}
\xi_1\\
\xi_2\\
\xi_3\\
\xi_4
\end{bmatrix}
\equiv
\int dk_z dk_T \frac{e^{i (z-u t) k_z} \Omega_1(k) }{k D_L(k)}
\begin{bmatrix}
J_0(\rho k_T)k_T k^2 \\
J_0(\rho k_T) k_T k_z^2\\
J_0(\rho k_T) k_T^3\\
J_1(\rho k_T) k_z k_T^2
\end{bmatrix}
\end{equation}
\begin{equation}
\begin{bmatrix}
\xi_5\\
\xi_6
\end{bmatrix}
\equiv
\int dk_z dk_T \frac{e^{i (z-u t) k_z} \Omega_2(k) }{k^2 D_T(k)}
\begin{bmatrix}
J_0(\rho k_T)k_z k_T^3\\
J_1(\rho k_T) k_T^2 k_z^2
\end{bmatrix}
\end{equation}
\begin{equation}
\begin{bmatrix}
\xi_7\\
\xi_8
\end{bmatrix}
\equiv
\int dk_z dk_T \frac{e^{i (z-u t) k_z}}{D_L(k)}
\begin{bmatrix}
J_1(\rho k_T)k_T^2\\
J_0(\rho k_T) k_z k_T
\end{bmatrix}
\end{equation}
\begin{equation}
\begin{bmatrix}
\xi_9\\
\xi_{10}\\
\xi_{11}\\
\xi_{12}
\end{bmatrix}
\equiv
\int dk_z dk_T \frac{e^{i (z-u t) k_z}}{k^2 D_T(k)}
\begin{bmatrix}
J_1(\rho k_T) k_T^2 k^2\\
J_1(\rho k_T) k_T^4\\
J_1(\rho k_T) k_T^2 k_z^2\\
J_0(\rho k_T) k_z k_T^3
\end{bmatrix}
\end{equation}
where the $k_z$ integration runs from $\pm\infty$ and the $k_T$ integration ranges from $0\rightarrow\infty$.  Since each $\xi_i$ contains a double integration which ranges over an infinite interval, care must be taken to cast each term in a form that will tend to zero as quickly as possible in order to make numerical evaluation feasible.  In the case of the integrals over $d k_z$, we can take advantage by the exponential term $e^{i (z-u t) k_z}$ by {\it bending the contour} in the imaginary plane.  The strategy works as follows:  start with a function of the form
\begin{equation}\label{kzform}
\int_{-\infty}^{\infty}e^{i (z-u t) k_z} f(k_z) d k_z
\end{equation}
where $f(k_z)$ is an analytic function.  Next, bend the contour of the integral in the complex plane by making the variable change
\begin{equation}
k_z \rightarrow r e^{i \alpha \text{Sign}[r]}
\end{equation}
and
\begin{equation}
d k_z \rightarrow dr e^{i \alpha \text{Sign}[r]}
\end{equation}
where the limits of integration on $r$ are from $-\infty$ to $\infty$.  Under this change, (\ref{kzform}) becomes
\begin{equation}\label{rform}
\int_{-\infty}^{\infty}\exp{[i (z-u t) r e^{i \alpha \text{Sign}[r]}]} f(r e^{i \alpha \text{Sign}[r]}) e^{i \alpha \text{Sign}[r]} d r
\end{equation}
and now the exponential term is damped in (\ref{rform}) [provided $\alpha$ has the same sign as $(z - u t)$] making the integral more numerically manageable.  (\ref{rform}) and (\ref{kzform}) will be equivalent if the integrand vanishes as $r\rightarrow \pm \infty$ and no poles from $f(k_z)$ are crossed when bending the contour (this places a restriction on the value of $\alpha$).

Before applying this technique it is necessary to consider where the $\xi_i$ have poles.  Any poles located directly on the imaginary axis will not affect the result so we can ignore those.  That leaves us to consider where the zeros of $D_T(u k_z)$ and $D_L(u k_z)$ are.  An analytic solution for the zeros of $D_T(u k_z)$ and $D_L(u k_z)$ is impossible because of the logarithmic terms.  However, the zeros can be searched for numerically.  Using the parameters ${\bf u} = 0.99 \hat{z}$ and $m_{\rm D} = 0.4$ GeV, it is found that all poles are sufficiently close to the imaginary axis so that the value of $\alpha$ used is not restricted for any reasonable value for terms with $D_T(u k_z)$ in the denominator.  I choose to use $|\alpha| = 1$.  However, for terms with $D_L(u k_z)$ in the denominator, it is found that for values of $k_T \lesssim 1/2$ poles become increasingly close to the real axis.  For this reason, I choose to break up the integration in the terms with $D_L(u k_z)$ in the denominator into a term
\begin{equation}
\int_{1}^{\infty} d k_T \int_{-\infty}^{\infty} d k_z
\end{equation}
which I bend in the complex plane with a value of $|\alpha| = 1$ and a term
\begin{equation}
\int_{0}^{1} d k_T \int_{-\infty}^{\infty} d k_z
\end{equation}
which I do not bend in the complex plane.  For the terms I do not bend in the complex plane it is usually necessary to subtract off the asymptotic form of the $k_z$ integration in order to make the numerical integration efficient.  The asymptotic form must then be added back.  Similarly, for the terms I do bend in the complex plane, I usually must subtract off the asymptotic form of the $k_T$ integration to provide efficient convergence.

In the case of $z = u t$, it does not make sense to bend the $k_z$ integral in the complex plane since there will be no damping term from the exponential.  In this case, I instead perform the numerical integration by subtracting off the unscreened form of $\xi_i$ in the integrand (which has the same large $k$ form) and then re-adding after evaluation.

To make these ideas more concrete, consider $\xi_7$ for the case of $z \neq u t$.  I write
\begin{equation}
\int_{-\infty}^{\infty} d k_z \int_{1}^{\infty} d k_T e^{i (z-u t) k_z} J_1(\rho k_T) \left(\frac{k_T^2}{D_L(u k_z)} - 1\right)
\end{equation}
where I have subtracted 1 to make the $k_T$ integration more efficient.  I bend the above contour in the complex plane and perform the integration and then re-add
\begin{equation}
\int_{-\infty}^{\infty} d k_z \int_{1}^{\infty} d k_T e^{i (z-u t) k_z} J_1(\rho k_T) \sim \delta(z-u t)
\end{equation}
which can be ignored, since it only contributes when $z = u t$.  Next, I integrate
\begin{equation}
\begin{split}
&\int_{-1}^{1} d k_z \int_{0}^{1} d k_T e^{i (z-u t) k_z} J_1(\rho k_T) \frac{k_T^2}{D_L(u k_z)}\\
& + \int_{1}^{\infty} d k_z \int_{0}^{1} d k_T e^{i (z-u t) k_z} J_1(\rho k_T) \left(\frac{k_T^2}{D_L(u k_z)} - \frac{k_T^2}{k_z^2}\right)\\
& + \int_{-1}^{-\infty} d k_z \int_{0}^{1} d k_T e^{i (z-u t) k_z} J_1(\rho k_T) \left(\frac{k_T^2}{D_L(u k_z)} - \frac{k_T^2}{k_z^2}\right)
\end{split}
\end{equation}
where I have subtracted $k_T^2/k_z^2$ to make the $k_z$ integration more efficient.  I then must re-add
\begin{equation}
\begin{split}
2 \text{Re}\left[\int_{1}^{\infty} d k_z \int_{0}^{1} d k_T e^{i (z-u t) k_z} J_1(\rho k_T)\frac{k_T^2}{k_z^2}\right] = \\
\frac{J_2(\rho)}{\rho}\left(2 (\cos (z)+z \text{Si}(z))-\pi  |z|\right).
\end{split}
\end{equation}
Considering $\xi_7$ for the case of $z = u t$, I write
\begin{equation}
\int_{-\infty}^{\infty} d k_z \int_{0}^{\infty} d k_T J_1(\rho k_T) \left(\frac{k_T^2}{D_L(u k_z)} - \frac{k_T^2}{k_T^2 + k_z^2}\right)
\end{equation}
and then analytically re-add [see (\ref{nascent})]
\begin{equation}
\begin{split}
\int_{-\infty}^{\infty} d k_z \int_{0}^{\infty} & d k_T e^{i (z-u t) k_z} J_1(\rho k_T)\frac{k_T^2}{k_T^2 + k_z^2} \\
&=\frac{\pi \rho}{(\rho^2 + (z-u t)^2)^{3/2}}
\end{split}
\end{equation}
and evaluate at $z = u t$.

\end{document}